# Forensics for Microsoft Teams

Marco Nicoletti, Massimo Bernaschi

**ABSTRACT**

Microsoft Teams is a collaboration and communication platform developed by Microsoft that replaces and extends Microsoft Skype for Business. It differs from Skype for Business by the fact that it exists only as part of the Microsoft 365 products whereas Skype for Business can be deployed completely or partly on premise.

During the pandemic emergency in 2020 and 2021 Microsoft Teams has increased dramatically its base of users as most of the meetings and the communications had to be conducted in virtual environments by users working remotely. Microsoft Teams allows users to collaborate sending and sharing information virtually with anyone internal or external to the organization with PCs and mobile devices, therefore it requires a careful review of all the security configurations and procedures within the organization.

Microsoft Teams infrastructure can also be integrated with PSTN telephone services, natively within the Microsoft 365 services or by integrating other

PSTN service providers. Therefore, its architecture extends the perimeter that could maliciously be exploited for an attack.

Microsoft Teams features can also be extended by Apps. There are hundreds of Apps developed by Microsoft and by other companies to address the various needs of modern collaboration. "Walkie Talkie", one of those Apps, transforms the Teams client installed in a mobile phone in a Walkie Talkie communication device for registered users.

The goal of this paper is to describe different Teams' usage scenarios and to analyse Teams' PSTN and Teams' Walkie Talkie communication scenarios describing forensics procedures to investigate inappropriate users' conducts.

**Keywords**: Microsoft Teams, VoIP Forensics, Walkie Talkie

# 1. INTRODUCTION

Microsoft Teams is a very powerful communication tool whose services can be grouped into three main categories provided to both PCs and mobile devices:

1. **Services for Meetings**: such as the creation of Teams and Channels. Microsoft Sharepoint and Microsoft OneDrive for Business are used to store the conversations and the files, to allow users' co-authoring and to record file versioning. Microsoft Exchange is integrated to allow sending email messages to Channels, supporting the integration with Calendar, Tasks and free-busy times. The Microsoft Teams Live Event allows events up to 10.000 users, the Artificial Intelligence Cloud services allow real time speech-to-text transcription, and the translations in several languages. This category of services includes everything required to prepare, to conduct and to follow up a meeting making it efficient and productive.
2. **Services for Real Time Communications**: supports VoIP Communications with other users internal or external to the organization, Audioconferencing, Videoconferencing, and Data Sharing in real time mode. They include also the Teams PBX services such as Call Queues and Auto attendants, the voice routing policies, and the architectures to integrate PSTN phone services.
3. **Custom services provided by Apps**: Teams can extend its functionalities in a wide set of different scenarios by integrating external Apps provided by Microsoft and other companies. The Walkie Talkie App is an example in this category.

Microsoft Teams exists only as part of the Microsoft 365 products, and within the Microsoft 365 environment it inherits all the security policies to ensure the Access Control and Identity Protection, the Threat Protection, the Information Protection, and the Security Management. Within the Microsoft 365 environment, the Microsoft Teams specific features can be configured by **Settings** and by **Policies**: the settings determine how a feature is globally provided whereas a policy is a collection of settings that can be assigned to users or group of users.

Microsoft Teams can be managed by the Microsoft Teams Admin Center, which provides a set of Analytics Usage Reports (Figure 1) and a very detailed report for each VoIP call including information like the microphone that has been used (Figure 2), the network parameters in both directions during the conversation (Figure 3), and other information useful to address specific forensic questions[1]. Figure 2 and Figure 3 specifically show analytic data about a PSTN phone call from a PC Teams client to a mobile phone: this scenario is going to be more and more common as Smart work and Modern work require users to make office phone calls from home or from remote locations as if they were from their office.

---

[1] All names shown in all the figures of this article are invented or belong to people who have provided their explicit approval for possible publication.

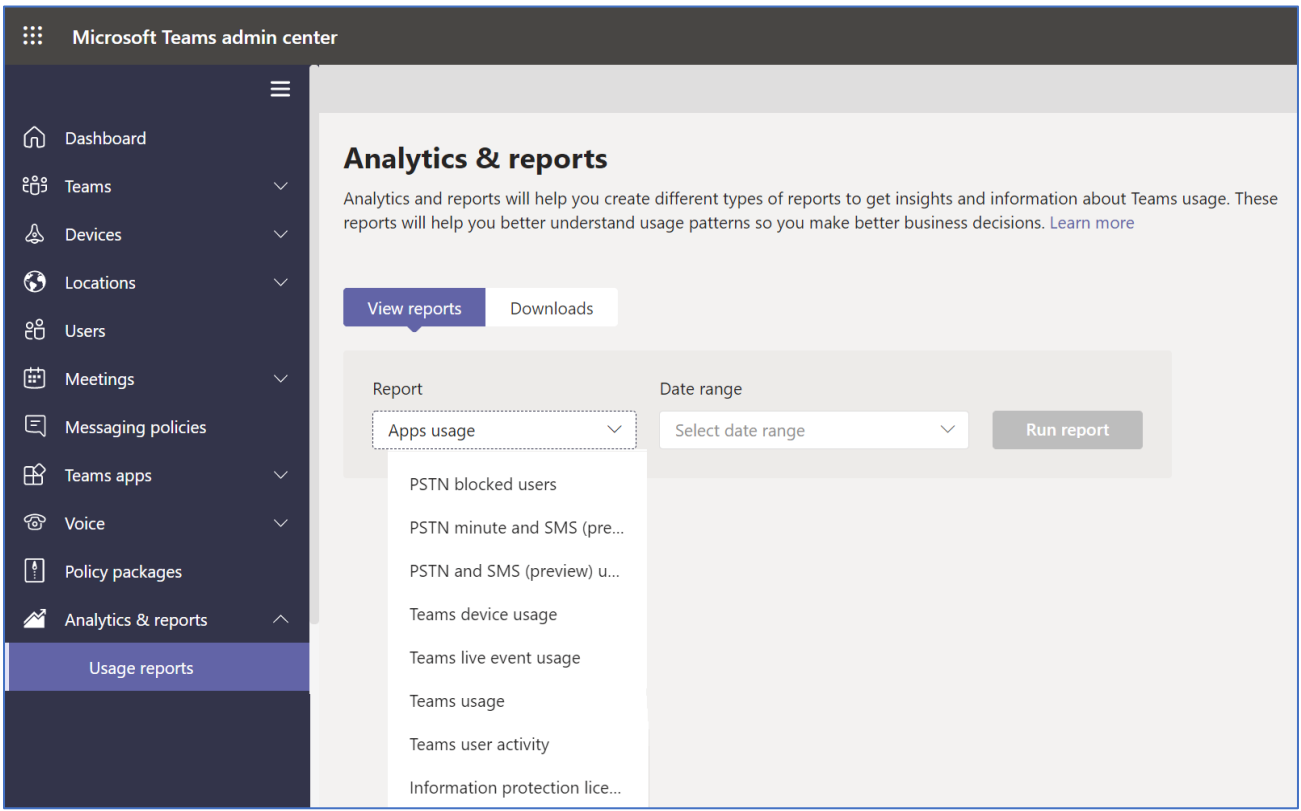

*Figure 1: Analytics Usage Reports available in the Microsoft Teams admin center.*

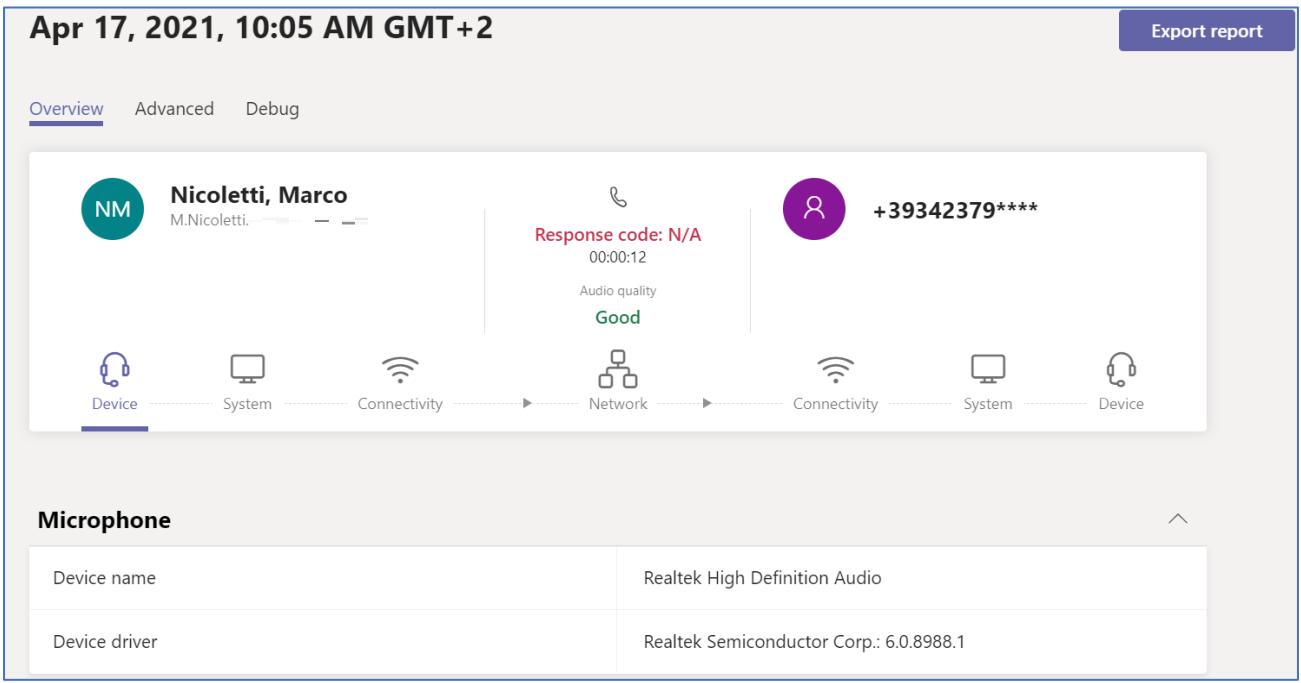

*Figure 2: Specific call details collected during a VoIP PSTN conversation.*

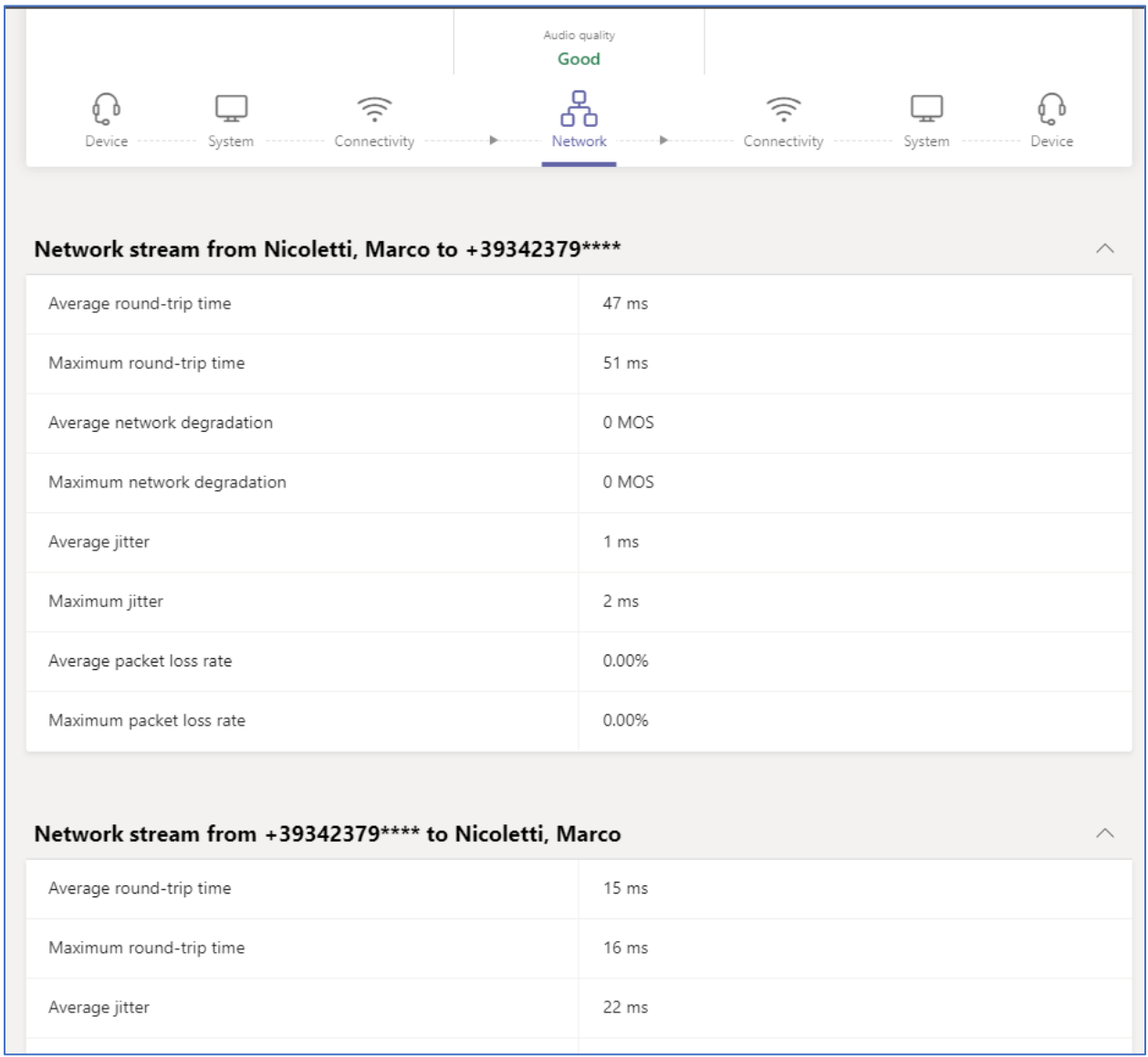

*Figure 3: Network parameters collected during a VoIP PSTN conversation.*

The Microsoft Teams Admin Center also provides a Main Dashboard to monitor the activities and a Call Quality Dashboard to analyse the quality of the traffic in each subnet of the network, provided that the network topology had been uploaded in the Tenant. The same information can be viewed also through a Business Intelligence Desktop client (Power BI) as shown in Figure 4.

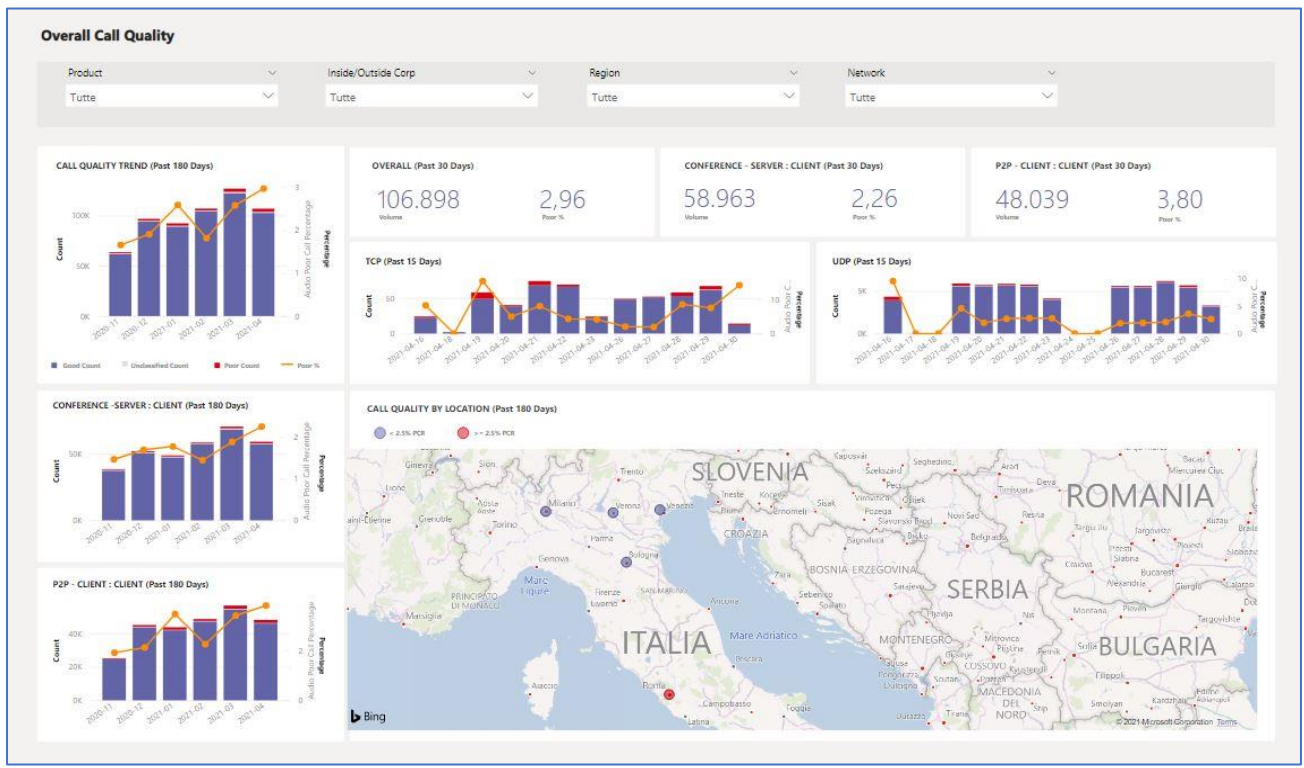

*Figure 4: Call quality information accessed through Power BI.*

Finally, the Microsoft Teams Admin Center allows the administrators to define which Apps can be used within Microsoft Teams.

In the present work we describe how to carry out an investigation about common Microsoft Teams activities. In particular, the paper aims at answering the following sample questions that may arise in a forensics investigation dealing with a Teams' VoIP scenario:

1. What is the average usage of Teams services made by a sample population of 1.000 users? What statistical analysis can be made about the Teams' usage of the sample population that may help to control the communications and prevent security breaches?
2. Someone reported an unsolicited communication about bullying or other types of crimes. Where and how can the investigator collect evidence about what really happened?

3. What are the risks of integrating the IP data network of a Microsoft Teams' environment with the external PSTN phone network? Where and how can the investigator collect evidence about what happened in a Teams-PSTN integrated environment?
4. Can we consider Teams Walkie Talkie communications via IP safer than traditional on-air radio Walkie Talkie communication? Is there resilient information in the Teams' environment that can help an investigator to trace a Teams' Walkie Talkie communication?

In order to answer the first two questions, we will use the Microsoft 365 Security and Compliance functionalities and the Microsoft Teams admin center looking for usage statistics and for evidence about unsolicited VoIP communications that had at least one user of the tenant as a participant.

In order to answer our third question, we will investigate a Microsoft Teams' environment integrated with the external PSTN by a certified AudioCodes Session Border Controller (SBC) using the Microsoft Teams admin center functionalities, a Wireshark software installed in a Windows Azure Virtual Server to which we redirect the SBC traffic, and a Syslog Viewer connected to the SBC.

In order to answer our fourth question, we will set up a Lab for Walkie Talkie and we will trace the network communication to/from a mobile device where it is installed a Walkie Talkie client.

The rest of the paper is organized as follows: Section 2 provides a high-level description of the Microsoft Teams architecture; Section 3 describes some basic forensics activities whereas Section 4 and 5 describe in detail the steps of the forensics analysis of a Teams' PSTN call and of a Walkie-Talkie call, respectively; Section 6 concludes the work providing a summary of the main findings and suggest the next steps.

## 2. MICROSOFT TEAMS ARCHITECTURE

Microsoft Teams' architecture is provided only On-Cloud[2], not necessarily integrated with PSTN phone services. Additionally, Microsoft supports a native telephone PSTN connectivity by purchasing Calling Plans in Microsoft 365. Instead of purchasing the Microsoft Calling Plans, Microsoft Teams can be connected directly to a PSTN service provider, for example by a SIP Trunking solution.

Microsoft Teams can also be integrated with the PBX that already provides the phone services to the company or organization. There are two possible strategies for connecting with a local PBX: 1. "Downstream", where the local PBX remains connected to the PSTN provider and must be configured to route the calls to the Teams environment. 2. "Upstream", where a Teams certified Gateway/SBC is placed between the PSTN provider and the local PBX, and no additional configuration is required in the local PBX.

Figure 5 shows the Downstream architecture with the SBC deployed in Azure that we have used to test forensics techniques for Teams in the present work. Examples of the other Teams' architectures are discussed in Appendix A.

---

[2] There are architectures which include local devices for scenarios that plan to produce a significant amount of media traffic and want to optimize it on the premise network.

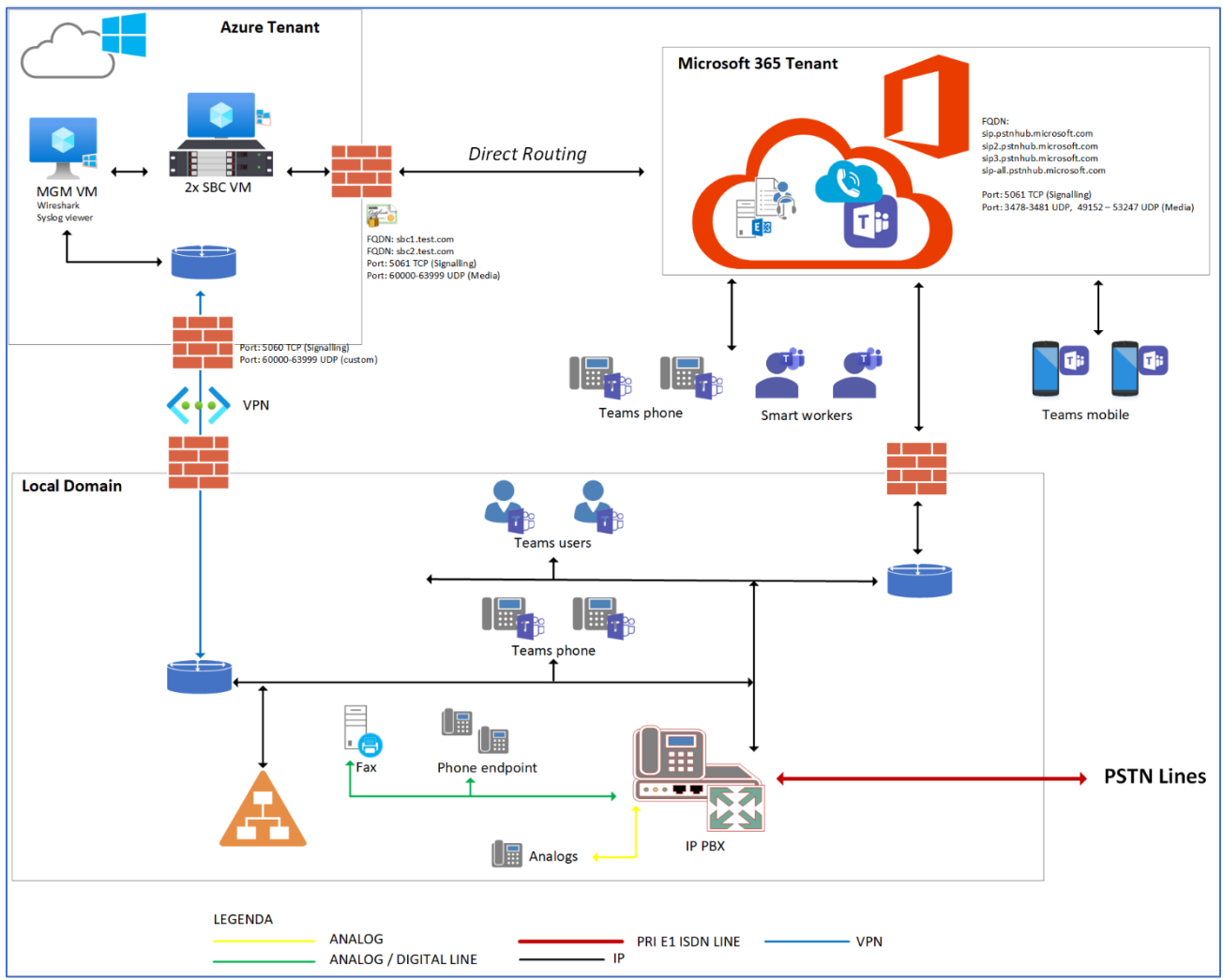

*Figure 5: Direct Routing downstream architecture with SBC(s) deployed in Azure*

As to the signalling, Teams uses the REST protocol within the Teams infrastructure and the Session Initiation Protocol (SIP) protocol to connect to external environments like the SBC. For Media traffic, Teams uses the Real-time Transport Protocol (RTP) preferably through the User Datagram Protocol (UDP), but if UDP ports are unavailable, then Teams tries the RTP via Transport Layer Protocol (TCP). The RTP flow is secured using Secure Real-time Transport Protocol (SRTP) in which only the payload is encrypted (Carolyn Rowe and others, 2018).

As to the Codecs, the Teams' Media traffic in the leg between the Teams' Cloud Media Processor and the SBC uses SILK, G.711, G.722 and G.729. It is possible to configure Teams also in "Media Bypass" mode, in this scenario the media traffic flows directly between the Teams' client and the SBC instead of flowing from Teams' client to Office 365 and then to SBC. The Media Bypass mode uses the same Codecs (Carolyn Rowe and others, 2021).

### a. Teams and Exchange, Sharepoint, Onedrive

Microsoft Teams depends on Sharepoint, Exchange and Onedrive. Whenever a user creates a Team to collaborate, it automatically creates a Sharepoint area that stores all the files that are shared in the Channel conversations of that Team. Similarly, whenever a user creates a Channel (a channel called "General" is created by default every time a user creates a Team), it automatically creates an email account in Exchange so that everyone can send an email directly to that channel.

As to Onedrive, *"For every user, the OneDrive folder Microsoft Teams Chat Files is used to store all files shared within private chats with other users (1:1 or 1:many), with permissions configured automatically to restrict access to the intended user only."* (Mike Plumley and others, 2021)

# 3. TEAMS FORENSICS

We start with an overview about the average usage of Teams services made by a sample population of 1.000 Microsoft Office users during the first three months of 2021 in a medium pandemic scenario when most of the users have been collaborating and communicating staying at home.

We may get the Teams' usage information from the "**Usage Reports**" in the "Teams admin center" selecting the reports for "Teams user activity", "Teams device usage" and "PSTN and SMS (preview) usage" (Figure 6).

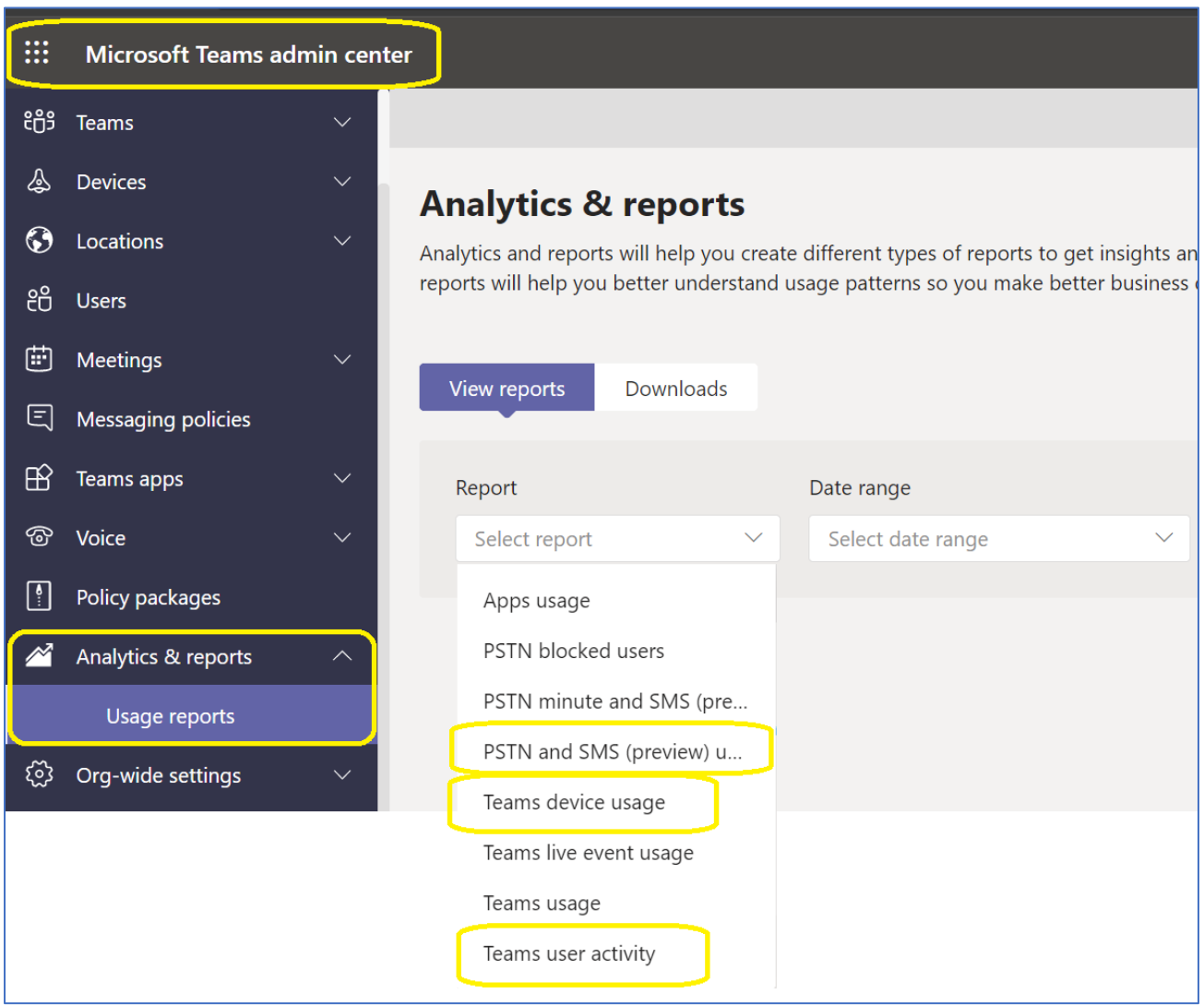

*Figure 6: Usage Reports in the Teams admin center*

Specifically, for "PSTN and SMS (preview) usage" we have selected the data about "Direct Routing" since this was the case study scenario that we are investigating.

The Microsoft Teams admin center allows us to export the reports to Excel, and this is the option we have selected. Table 1 shows a high-level summary of the data extracted.

| Usage | Last 7 days | Last 30 days | Last 90 days |
|---|---|---|---|
| **Total users of Teams services** | 1.171 | 1.224 | 1.272 |
| **1:1 Calls** | 36.368 | 139.960 | 419.001 |
| **Total Audio time** | 570 days 23 hours 12 minutes | 2251 days 7 hours 6 minutes | 6882 days 4 hours 2 minutes |
| **Total Video time** | 225 days 9 hours 56 minutes | 881 days 1 hours 56 minutes | 2587 days 12 hours 1 minutes |
| **Average audio time per user** | 12 | 44 | 130 |
| **Average video time per user** | 5 | 17 | 49 |
| **Windows PC users** | 1.141 | 1.183 | 1.222 |
| **Mac users** | 4 | 9 | 13 |
| **iOS Phone users** | 247 | 270 | 294 |
| **Android Phone users** | 252 | 280 | 304 |
| **Linux users** | 0 | 0 | 1 |
| **Web browser users** | 15 | 64 | 139 |
| **Total number Teams PSTN calls** | 6.536 | 28.706 | 96.394 |
| **Total time Teams PSTN calls** | 12 days 1 hours 12 minutes | 53 days 21 hours 49 minutes | 180 days 4 hours 49 minutes |

*Table 1: Teams' usage data*

Table 2 shows a drill down on data extracted from the Microsoft Teams Admin Center with the timings of the 10 users more active according to their usage of Audio and Video time during the last 7 days. It is apparent that there is a surprisingly large fraction of time dedicated by users for communicating with Teams and this is what we must be ready to investigate with a forensic methodology.

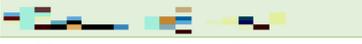

*Table 2: Sample of a drill down on Teams' usage data in the last 7 days*

The **Microsoft Call Quality Dashboard** in the Microsoft 365 admin center stores additional information about the usage of Teams. It is designed to present the Teams' Overall Call Quality and it can provide the call quality data measured in each network segment[3]. This information might help if the forensic investigator should identify an activity that somehow temporarily compromised the performance of a sub network.

Teams stores the contents of the users (chat, files…) in Sharepoint, One Drive and Exchange according to the type of content. The knowledge of the integration between Teams and the other above-mentioned services allows the IT administrators to create the policies to maintain, delete, monitor, and search the content.

The **Content search in the Microsoft 365 compliance center** (Figure 7) allows for searching and understanding when a specific Teams' content was last accessed and by whom.

---

[3] To get the call quality data split by network segment, the network topology must be uploaded in the Tenant.

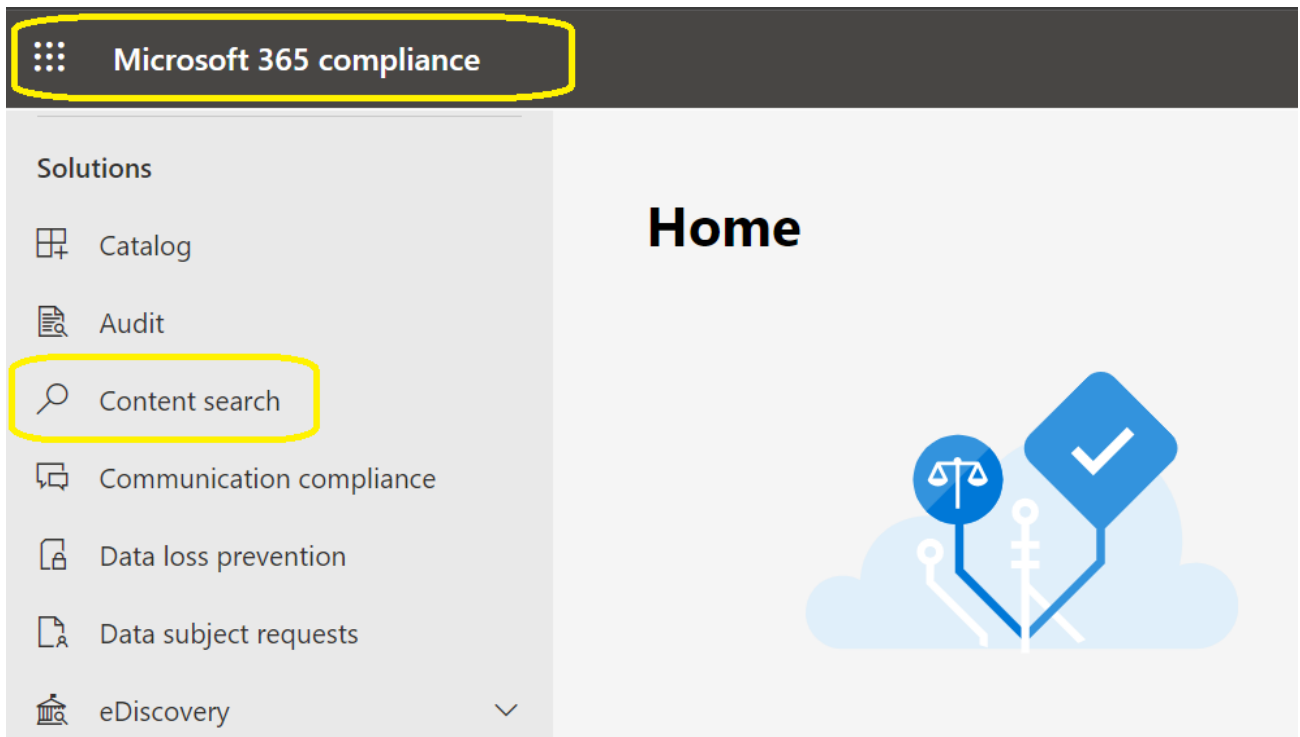

*Figure 7: Content search in the Microsoft 365 compliance center*

Let's create a search condition as in Figure 8, then write the keyword in a Private chat conversation and in a Channel conversation as in Figure 9 and Figure 10. We may now run our content search and find an exportable report about who has ever written that keyword and where (Figure 11).

In order to understand the results shown in Figure 11 let's remind that "*For a cloud-based user, Teams chat data (also called 1x1 or 1xN chats) is saved to their primary cloud-based mailbox.*" (Mark Johnson and others, 2021) and "*Records for messages sent in a private channel are delivered to the mailbox of all private channel members, rather than to a group mailbox. The titles of the records are formatted to indicate which private channel they were sent from.*" (Mark Johnson and others, 2021).

Similarly, we may use the **Advanced eDiscovery in the Microsoft 365 compliance center** that is *"…the electronic aspect of identifying, collecting and producing electronically stored information (ESI) in response to a request for production in a lawsuit or investigation. Capabilities include case management, preservation, search, analysis, and export of Teams data. This includes chat, messaging and files, meeting and call summaries. For Teams meetings and Calls, a summary of the events that happened in the meeting and call are created and made available in eDiscovery."* (Laura Williams and others, 2021). For the private nature of the data that can be discovered, using the eDiscovery functionalities requires appropriate permissions or roles that can run and export the findings. "*The primary eDiscovery-related role group in Security & Compliance Center is called eDiscovery Manager. There are two subgroups within this role group. eDiscovery Managers… [and] …eDiscovery Administrators*". (Mark Johnson and others, 2021)

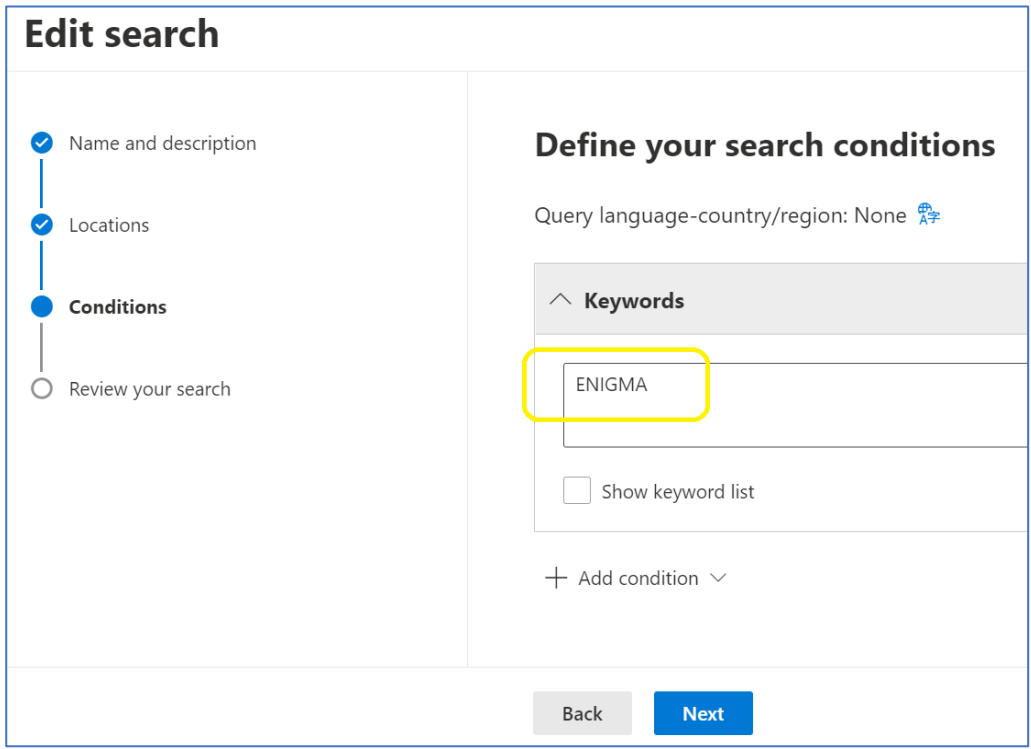

*Figure 8: Search condition*

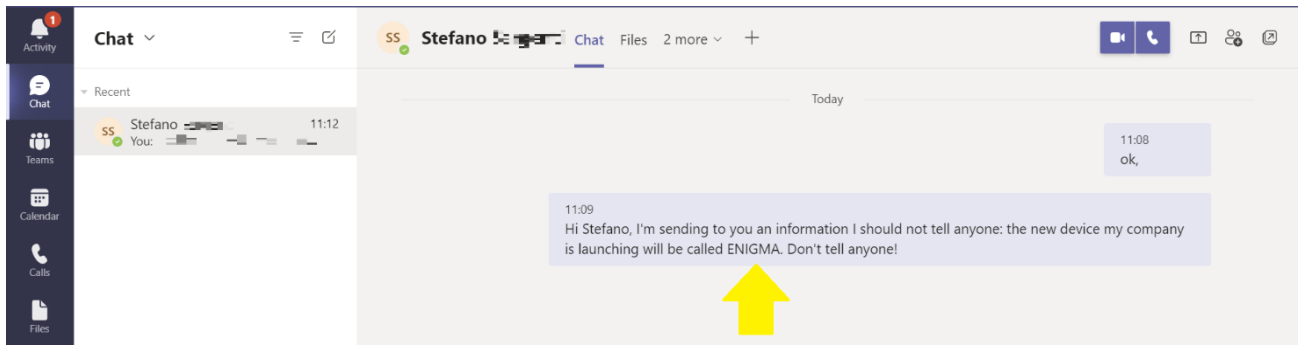

*Figure 9: Teams private chat*

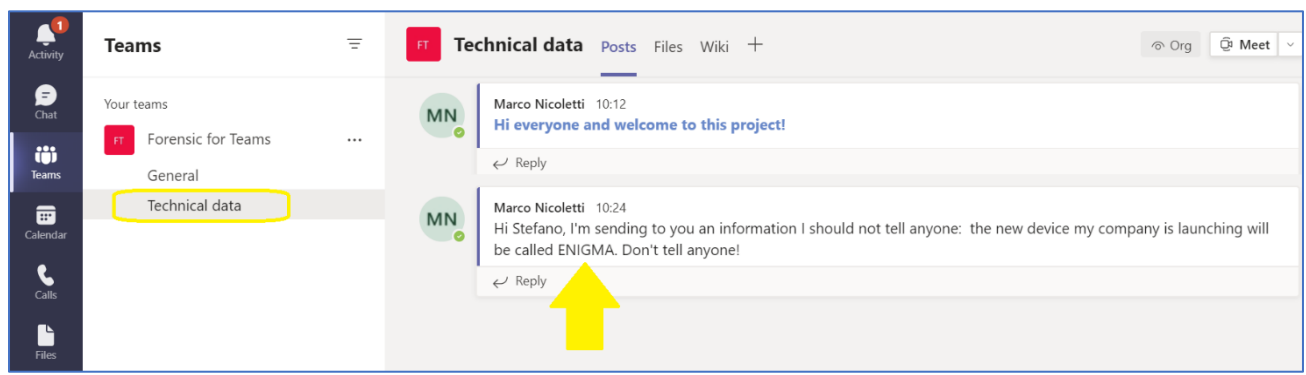

*Figure 10: Teams channel conversation*

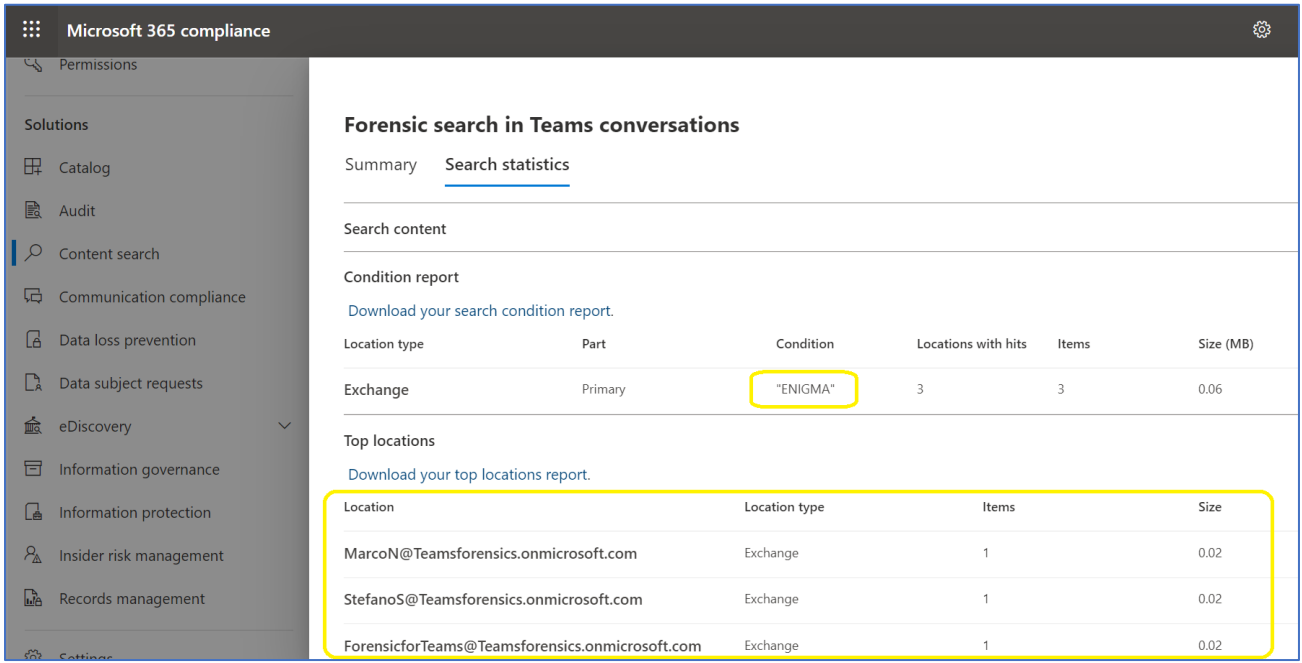

*Figure 11: Content search report*

Let's now mention the **Legal Hold**. *"When a reasonable expectation of litigation exists, organizations are required to preserve electronically stored information (ESI), including Teams chat messages that are relevant to the case. Organizations may need to preserve all messages related to a specific topic or for certain individuals… …Within Microsoft Teams, an entire team or select users can be put on hold or legal hold. Doing that will make sure that all messages that were exchanged in those teams (including private channels) or messages exchanged by those individuals are discoverable by the organization's compliance managers or Teams Admins."* (Mark Johnson and others, 2021).

The Table 3 lists the location where the Teams content is stored to help planning what locations submit to Legal Hold services.

| Scenario | Content location |
|---|---|
| Teams chats for a user (for example, 1:1 chats, 1:N group chats, and private channel conversations) | User mailbox. |
| Teams channel chats (excluding private channels) | Group mailbox used for the team. |
| Teams file content (for example, Wiki content and files) | SharePoint site used by the team. |
| Teams private channel files | Dedicated SharePoint site for private channels. |
| User's private content | The user's OneDrive for Business account. |
| Card content in chats | User mailbox for 1:1 chats, 1:N group chats, and private channel conversations or group mailbox for card content in channel messages. |

*Table 3: Content locations to place on legal hold to preserve Team's content. (Mark Johnson and others, 2021)*

## 4. FORENSICS OF A TEAMS' CALL

Every Teams' VoIP call is tracked in Microsoft 365 and the details can be accessed by the user interface of the Microsoft Teams admin center, by querying the Tenant with Microsoft Power BI (Business Intelligence), or by other specific tools like the AudioCodes One Voice Operations Center (OVOC) (AudioCodes, 2021).

Most of corporate Teams' environments allow their users to connect with Skype Consumer users and it is useful to recall that Skype Consumer users might have registered providing a fake name, and they could have connected from a public unmonitored or insecure network for which they obtained an account without providing any genuine information.

Figure 12, Figure 13 and Figure 14 show where the forensics investigator can search for the details of a VoIP call received by the user Marco Nicoletti. We see detailed information such as the drivers of microphone, speaker, and Wi-Fi of the caller's device and the network parameters registered for both parties during the call.

*Figure 12: Sample of device parameters collected in the Microsoft Teams admin center.*

*Figure 13: Sample of connectivity parameters collected in the Microsoft Teams admin center.*

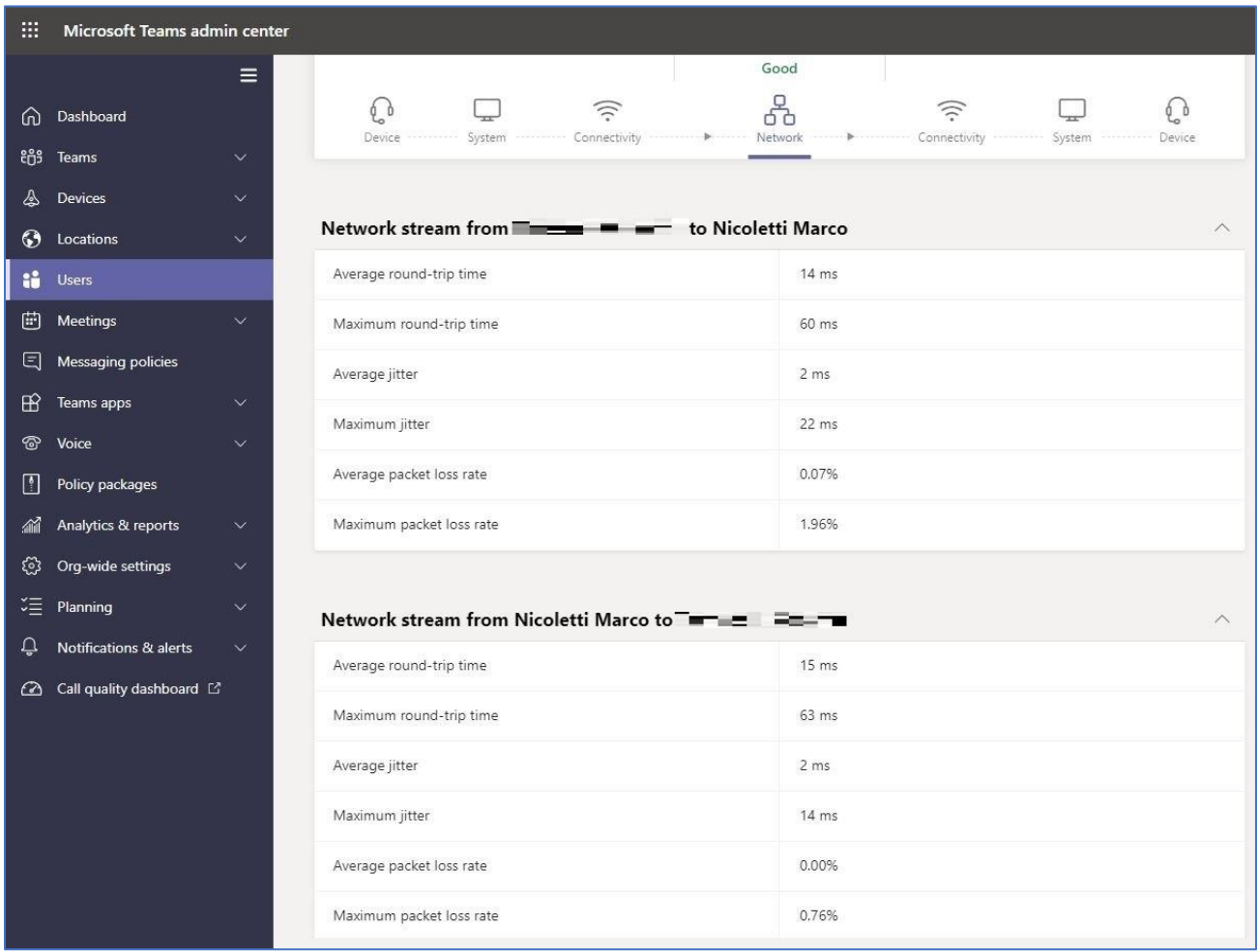

*Figure 14: Network information of a call.*

However, not all information is available in all scenarios: for example, the Wi-Fi driver description of the device of the caller is not available in the PSTN and Skype Consumer VoIP call scenarios. Table 4 shows the list of all the information stored in Microsoft 365 by Teams for both the caller and the recipient of the call.

| Communication Detail | Notes |
|---|---|
| Microphone device name | |
| Microphone device driver | |
| Speaker device name | |
| Speaker device driver | |
| System name (Computer name) | |
| Operating System | |
| Network connection Type | Separately available for VoIP and for Desktop Sharing |
| Wi-Fi driver description | Separately available for VoIP and for Desktop Sharing |
| Wi-Fi driver version | Separately available for VoIP and for Desktop Sharing |
| Wi-Fi signal strength | Separately available for VoIP and for Desktop Sharing |

*Table 4: Information available for a call.*

Whenever the investigator must collect information about a VoIP PSTN conversation from a Teams' Direct Routing Architecture (Figure 5) he/she could search for useful data also in the Session Border Controllers (SBC). The SBC acts as a gateway and a firewall for PSTN calls between Microsoft 365 and the PSTN network, therefore in the CDR history of the SBC each Inbound or Outbound call is shown subdivided into two trunks: one to Teams and the other one to the PSTN (the IP PBX in our scenario). Figure 15 shows an example of a Call-Detail Record (CDR) stored in a virtual AudioCodes SBC deployed in Azure: we may see information such as the phone number of the user who made the call, the phone number of the recipient of the call, the IP of the device that received the call in the PBX leg, the Office 365 IP that has received the call in Teams, the duration of the call and whether the call completed successfully, failed, received no answer, or terminated for other reasons. All this information can also be forwarded to a centrally managed CDR collector.

*Figure 15: Sample of SBC CDR.*

The SBC allows to collect also SIP logging. Figure 16 shows a sample of SIP logs collected during a phone call.

*Figure 16: SBC Logging.*

In the CDR of Figure 15 and in the SIP logs of Figure 16 we see remote connections with the IP 52.114.75.24 and 52.112.110.85, those are, as a matter of fact, the IPs published by Microsoft for Office 365 Teams in the ranges 52.112.0.0/14 and 52.120.0.0/14 (Carolyn Rowe and others, 2021).

SBC logging information for an environment with one thousand regular users average to about 150-200 Mbyte of uncompressed data (about 20Mbyte compressed) every day for every SBC in use[4]. Moreover, the logging is strictly chronological and therefore several calls overlap in the logging trace. Not all the software that is available to analyse SIP messages can open very large files. It is therefore important to plan, in advance, which procedure should be used in the event of a forensic investigation, considering specific software like OVOC (AudioCodes, 2021) or custom manipulation of text files to pre-select the parts of the logging files that are relevant for the investigation.

The SBC also provides CODEC transcoding features. This means that the SBC can decrypt the SRDP traffic from Teams to PSTN and vice versa and that it is possible to collect the RDP traffic in the SBC to listen to all the voice conversations of the users in both directions. Figure 17 shows an example of reproducing the audio call made by a Teams' client to a mobile phone on PSTN through the SBC. The blue wave initiates with the tone ringing, then continues with the audio sent by the PSTN mobile device, the dark grey wave reproduces the audio from the Teams' client.

---

[4] Usually in a production environment there are several SBC deployed for redundancy.

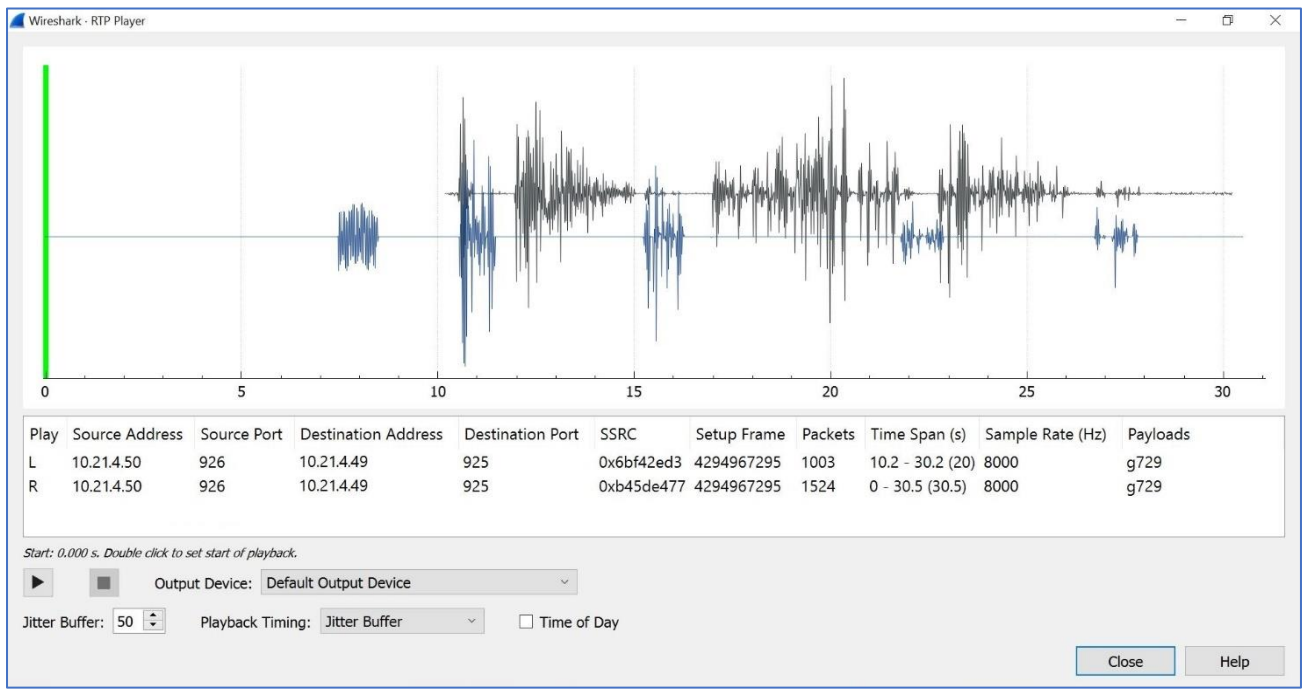

*Figure 17: Collecting the Audio of PSTN calls.*

In Appendix B we describe the detailed steps to configure the AudioCodes SBC to collect the CDR, the SIP Logs, and the Audio of VoIP calls.

## 5. WALKIE TALKIE

*"The Walkie Talkie app in Teams provides instant push-to-talk (PTT) communication… Walkie Talkie allows users to connect with their team using the same underlying channels they're members of. Only users who connect to Walkie Talkie in a channel become participants and can communicate with each other using push-to-talk, one at a time."* (Heidi Payne and others, 2021)

In short, the Walkie Talkie App simulates a radio device on Android, and it uses the Channels in Teams as if they were frequency channels to which the user could tune its device. Like any radio communication, only one user at a time can use the channel to speak, but every user that is connected to the channel can hear the audio conversations, exactly as if they were using a traditional walkie talkie radio. The difference is that the users of Walkie Talkie must authenticate the App in Office 365 offering the opportunity to the administrators to enforce the security and compliance controls we have so far discussed. Figure 18 shows an example of the Walkie Talkie App installed on an Android phone, connected to a Team's channel and ready to talk.

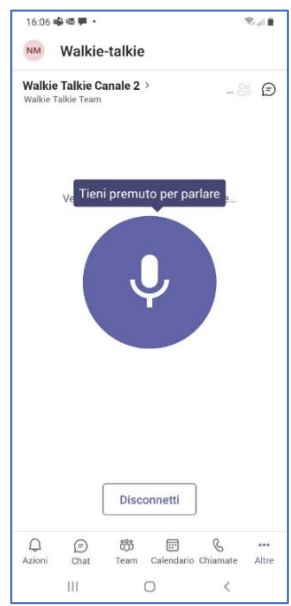

*Figure 18: Walkie Talkie app on a mobile device.*

We analysed the network traffic of a Walkie Talkie communication and we checked if the Microsoft Teams admin center functionalities could be used to discover the resilient information of Walkie Talkie communications.

The Android Wi-Fi network traces that we collected during our test show that the walkie talkie client has been communicating with several IPs in the range published by Microsoft for Office 365 Teams. Apart from the traffic to Microsoft for Office 365, during our test the Android device with the Walkie Talkie App made traffic only to Google (not relevant for our analysis) and with the local gateway, to which the Walkie Talkie App requested the DNS entry for "walkietalkie.teams.microsoft.com", the one we could see listed most of the times when we tracked the network traffic during a Walkie Talkie communication.

Figure 19 shows the network conversation of our Walkie Talkie App (IP 192.168.1.5) with the Microsoft for Office 365 IP range 52.112.0.0/14, among which the IP 52.114.74.99 of "walkietalkie.teams.microsoft.com".

| Address A | Port A | Address B | Port B | Packets | Bytes | Packets A → B | Bytes A → B | Packets B → A | Bytes B → A | Rel Start | Duration | Bits/s A → B | Bits/s B → A |
|---|---|---|---|---|---|---|---|---|---|---|---|---|---|
| 192.168.1.5 | 48851 | 52.114.104.172 | 443 | 5 | 683 | 3 | 433 | 2 | 250 | 0.727369 | 0.0779 | 44k | 25k |
| 192.168.1.5 | 38078 | 52.114.77.33 | 443 | 34 | 22k | 21 | 20k | 13 | 2538 | 0.945041 | 24.5994 | 6535 | 825 |
| 192.168.1.5 | 42429 | 52.114.74.99 | 443 | 7 | 579 | 4 | 354 | 3 | 225 | 8.592562 | 0.0777 | 36k | 23k |
| 192.168.1.5 | 42428 | 52.114.74.99 | 443 | 33 | 6029 | 15 | 2501 | 18 | 3528 | 8.633261 | 14.3282 | 1396 | 1969 |
| 192.168.1.5 | 37038 | 52.114.74.97 | 443 | 6 | 1244 | 4 | 620 | 2 | 624 | 8.637498 | 14.2181 | 348 | 351 |
| 192.168.1.5 | 42472 | 52.114.74.99 | 443 | 31 | 11k | 17 | 4186 | 14 | 7265 | 8.988481 | 13.7996 | 2426 | 4211 |
| 192.168.1.5 | 42473 | 52.114.74.99 | 443 | 27 | 11k | 14 | 3944 | 13 | 7214 | 22.991771 | 15.8236 | 1993 | 3647 |
| 192.168.1.5 | 46095 | 52.114.74.181 | 443 | 3 | 773 | 1 | 330 | 2 | 443 | 28.710377 | 0.1775 | 14k | 19k |
| 192.168.1.5 | 42433 | 52.114.74.211 | 443 | 3 | 273 | 2 | 172 | 1 | 101 | 30.795856 | 0.0316 | 43k | 25k |

*Figure 19: Office 365 IPs during a Walkie Talkie session*

We have also been tracking the network traffic in the scenario of two Android devices connected in the same local Wi-Fi network and communicating together with Walkie Talkie App. In such scenario we could not find any direct traffic to/from the two clients, which means that all communications had been routed by Office 365 Teams IPs to which both clients were registered. As expected, we found that there is no SIP traffic in the network.

We have also verified that the communications made by Walkie Talkie are not tracked in the Microsoft Teams admin center "Meetings and Calls" tab. Figure 20 shows an example of the listed conferences and calls made by the user Marco Nicoletti and there is no evidence of the Walkie Talkie conversations that the same user made in that timeframe.

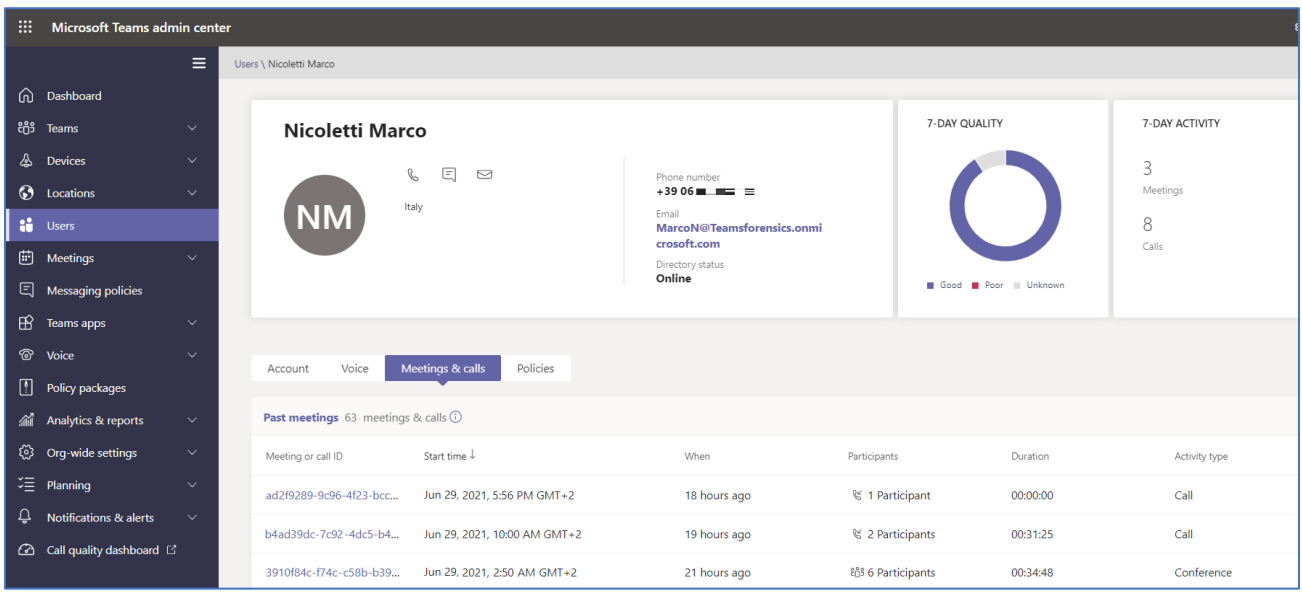

*Figure 20: Meeting & Call tracking tab for a Teams' user*

In Appendix C we describe the lab environment that we used to track the Wi-Fi network communication to and from the Android device with the Walkie Talkie app installed.

# 6. CONCLUSIONS AND FUTURE ACTIVITIES

All VoIP conversations made by a Teams' client to another Teams' client or by a Teams' client to/from a PSTN device are tracked in the Microsoft 365 and a forensics analysis can be made using the services provided by the Microsoft Teams admin center or by connecting to the Microsoft 365 Tenant with Power BI tools.

The SBC logging is strictly chronological, the size of the information logged is significant. It is important to plan, in advance, which procedure should be used in the event of a forensic investigation, considering specific software or custom manipulation of log files to pre-select what may be relevant for the investigation.

All IPs and Ports used by Microsoft Teams VoIP services are known, reserved and public.

When a Teams' environment is connected to the PSTN by a Direct Routing architecture with managed SBC(s), the VoIP conversation made by a Teams' client to/from a device in the PSTN network can be tracked in the SBC(s) by the SBC administrator. Moreover, in such scenario the VoIP audio of both parties can be listened by the SBC administrator, including the initial ringing tones. The VoIP Audio can be regularly monitored using Wireshark and stored to perform further analysis later. Such investigations would be completely transparent for the PSTN provider, meaning that the PSTN provider cannot understand that the conversation of his/her customer has been tracked and listened in the SBC gateway.

The privacy and the confidentiality risks of integrating the IP data network of a Microsoft Teams' environment with the PSTN phone network can be mitigated by arranging and enforcing the available Microsoft 365 compliance rules, such as enforcing an Audit Log to track the administrators' activity and allowing the connection to the SBC(s) only from a specific server in Azure.

As a best practice, we recommend configuring the SBC(s) so that they accept connections only by the IP of a management server in Azure, to which it is possible to authenticate only trough Azure Bastion (Cheryl McGuire and Others, 2021).

The conversations made by Walkie Talkie App cannot be analysed using the services provided by the Microsoft Teams admin center. In practice, we may track a Walkie Talkie communication only by collecting the network traffic.

Compared to the traditional Walkie Talkie radio solutions, the Walkie Talkie App provides a great degree of security because the client devices register to the Teams' services, and it can be monitored in accordance with the security and compliance settings configured for the Microsoft 365 Tenant.

a. **Next steps**

**Microsoft 365 Defender** portal is available for each Microsoft tenant[5] to group together different threat signals and to determine the full scope and impact of a threat. It consists of "…*a unified pre- and post-breach enterprise defense suite that natively coordinates detection, prevention, investigation, and response across endpoints, identities, email, and applications to provide integrated protection against sophisticated attacks.*" (jcaparas and others, 2021).

**Microsoft Defender for Endpoint** is one of the suites available within the Microsoft 365 Defender portal to collect and retain for a default time of 180 days information from configured devices in a customer tenant including file data (such as file names, sizes, and hashes), process data (running processes, hashes), registry data, network connection data (host IPs and ports), and device details (such as device identifiers, names, and the operating system version). (jcaparas and others, 2021).

Our future activities will consist in creating a methodology and tools to leverage all the information discussed in this article and those available in the Microsoft 365 Defender portal to the purpose of collecting a rich and consistent set of evidence useful for any VoIP forensic analysis involving Microsoft Teams.

---

[5] Specific licensing may be required.

# APPENDIX A: UPSTREAM AND DOWNSTREAM ARCHITECTURES

Figure 21 shows a scheme with a typical Teams architecture with no integration of the traditional PSTN system.

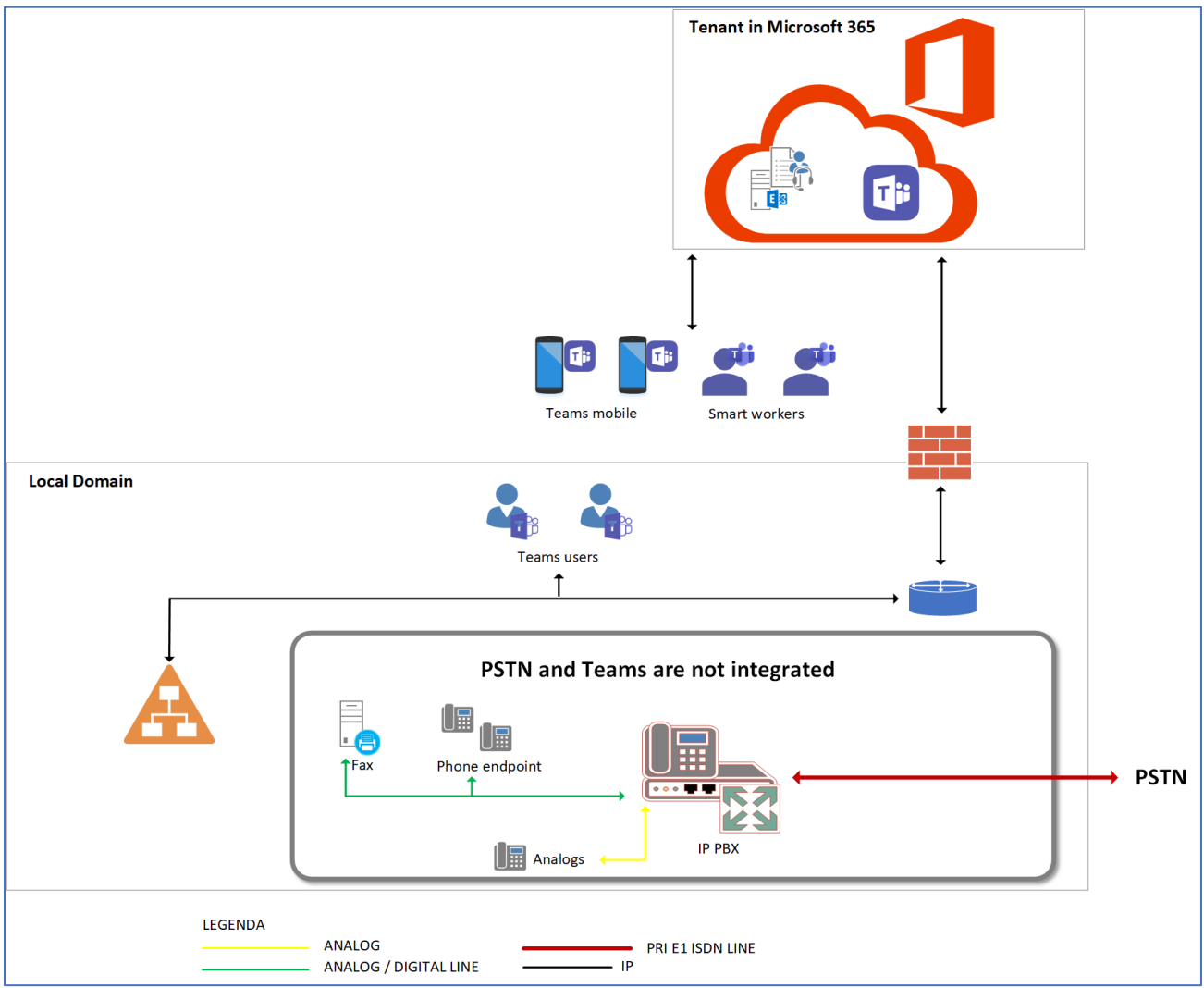

*Figure 21: Teams architecture without PSTN integration*

Figure 22 shows a scheme with a typical Teams architecture connected directly to a PSTN service provider.

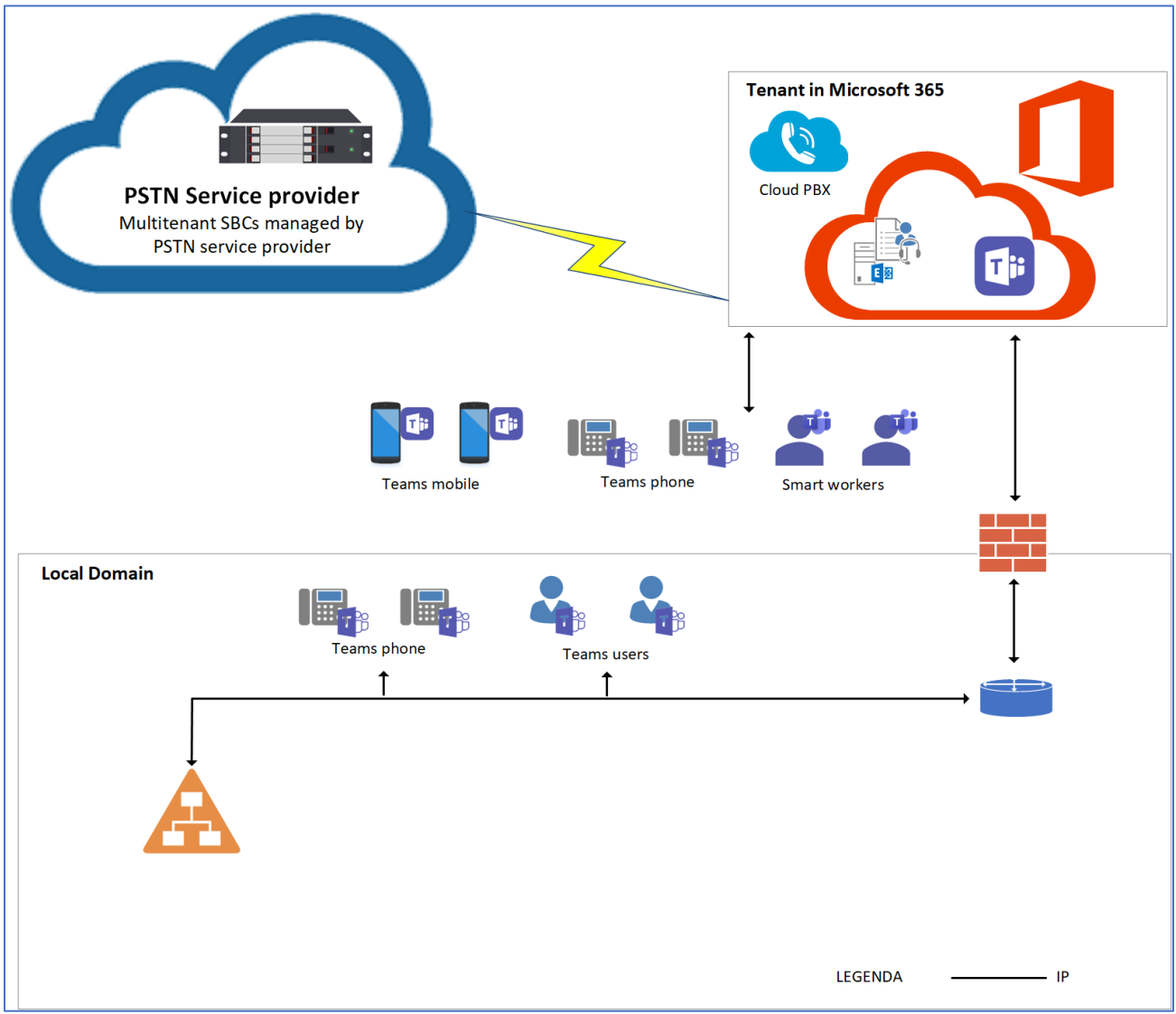

*Figure 22: Teams architecture with a PSTN service provider.*

Microsoft Teams can also be integrated with on-premise PBX:

1. "Downstream", where the local PBX remains connected to the PSTN provider;
2. "Upstream", where a Teams certified Gateway/SBC is placed between the PSTN provider and the local PBX.

The "Downstream" strategy offers different architectural choices, where the SBC can be virtual and placed in Azure or in an on-premises VM farm, or the SBC can be physical and placed on-premises. If the PBX has an IP interface and a license to allow IP traffic, then the connectivity between the Gateway/SBC and the PBX can be made via IP, otherwise when the Gateway/SBC is physical and placed on-premises, it is possible to connect the Gateway/SBC and the PBX using a physical BRI/PRI cable. This is a common scenario when there is an old PBX and it is not convenient to invest money to add IP connectivity on the PBX. Figure 23 shows the diagram using a BRI/PRI cable in an upstream architecture. In the "Upstream" scenario a physical component must be placed on premise to be linked to the BRI/PRI cables and/or to the PSTN, but most of the SBC services can still be provided by a virtual server placed in Azure or in a local VM farm.

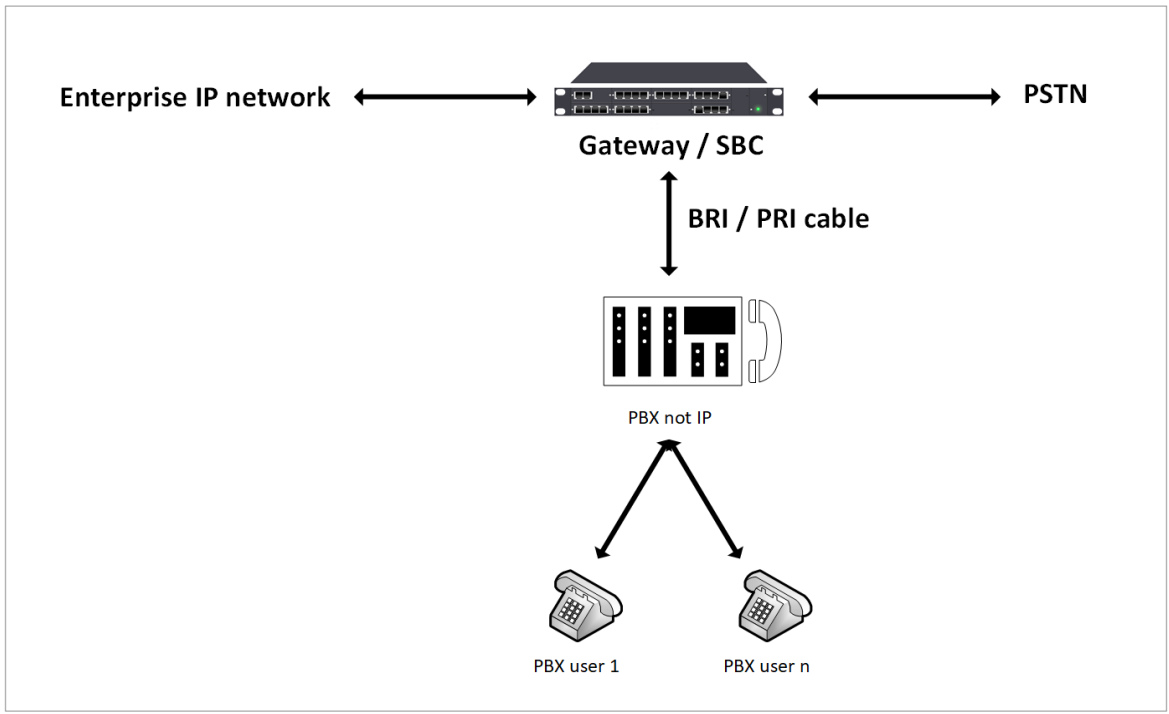

*Figure 23: Sample of upstream scenario with BRI/PRI cable*

# APPENDIX B – METHODOLOGY FOR COLLECTING FORENSICS EVIDENCE FROM SBC

This Appendix describes how to collect Call-Detail Record (CDR), SIP Signaling logs and the Audio of a VoIP call from a virtual AudioCodes SBC deployed in Azure. This procedure would be the same for any virtual or physical AudioCodes SBC.

To collect Call-Detail Record (CDR) select "Monitor" in the main menu option of the SBC and then "SBC CDR History" as shown in Figure 24 and you will be presented with the list of the Call-Detail Record (CDR) stored in the SBC[6]. This information can also be forwarded to a centrally managed CDR collector

---

[6] CDRs reset when you update firmware and reboot the SBC.

configuring the "SBC CDR Remote Servers" menu section as shown in Figure 25.

*Figure 24: SBC CDR History.*

*Figure 25: SBC CDR Remote Servers configuration.*

We can use the SBC also to track the SIP logging information. The SIP Logging can be collected real-time by a Syslog Viewer, or it can be centrally collected and managed for further analysis. AudioCodes provides a Syslog Viewer that can be freely downloaded from the AudioCodes Utilities[7]. For collecting the real time logs used in this article we used the version 1.58 shown in Figure 26, connected to the SBC selecting "File" and "Connect To…" as shown in Figure 27 and configuring the connection parameters as shown in Figure 28.

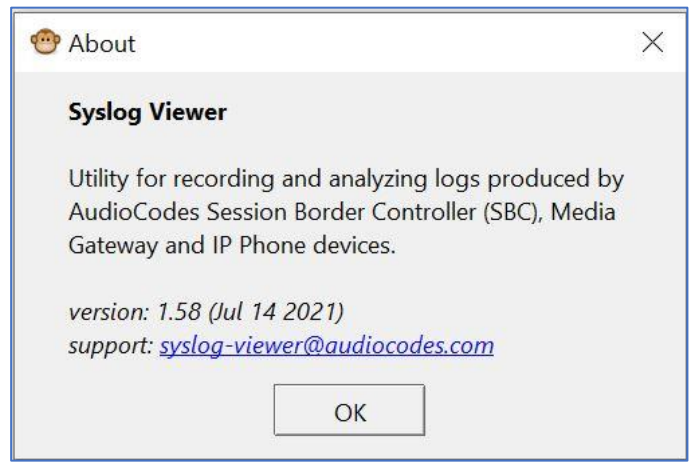

*Figure 26: AudioCodes Syslog Viewer.*

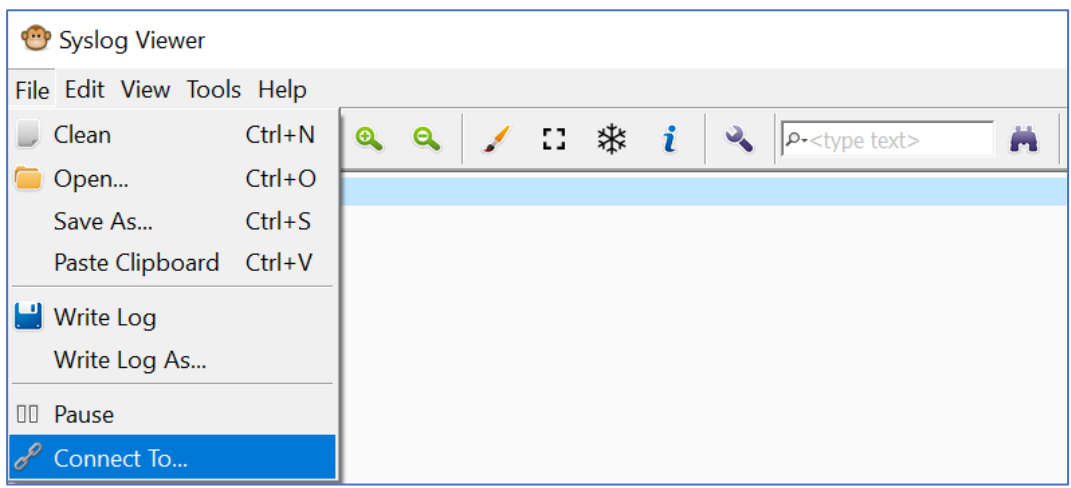

*Figure 27: Syslog viewer menu.*

---

[7] AudioCodes Utilities: http://redirect.audiocodes.com/install/index.html

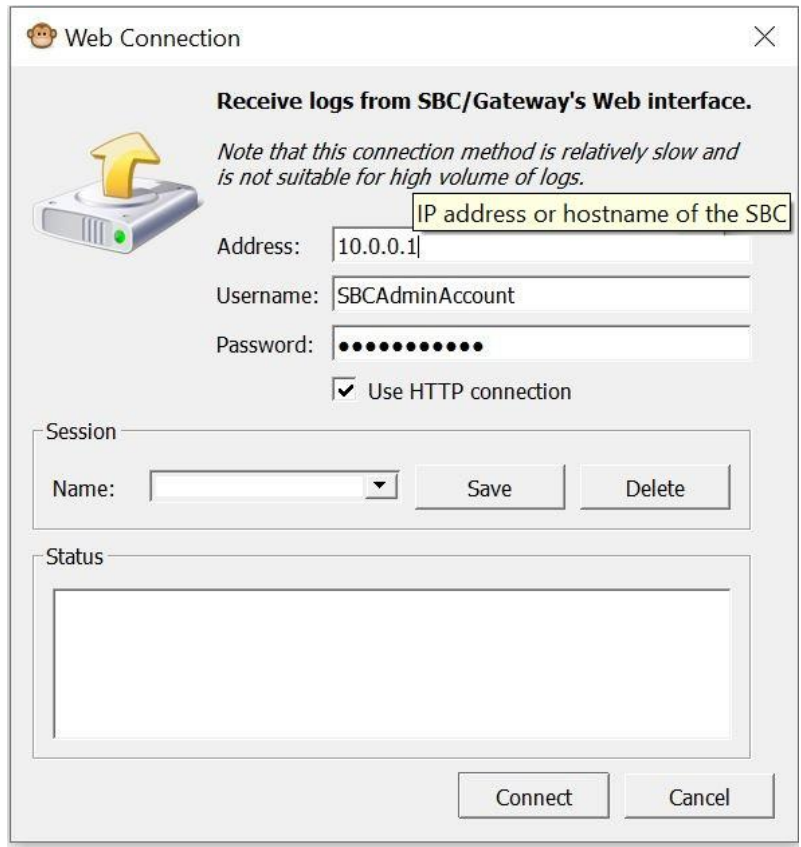

*Figure 28: Syslog viewer connection parameters.*

To configure the SBC to send the logs to a centrally managed server, we can select the section "Troubleshoot" of the SBC and configure the settings in the menu item "Logging settings" including the IP of the remote logging server as shown in Figure 29.

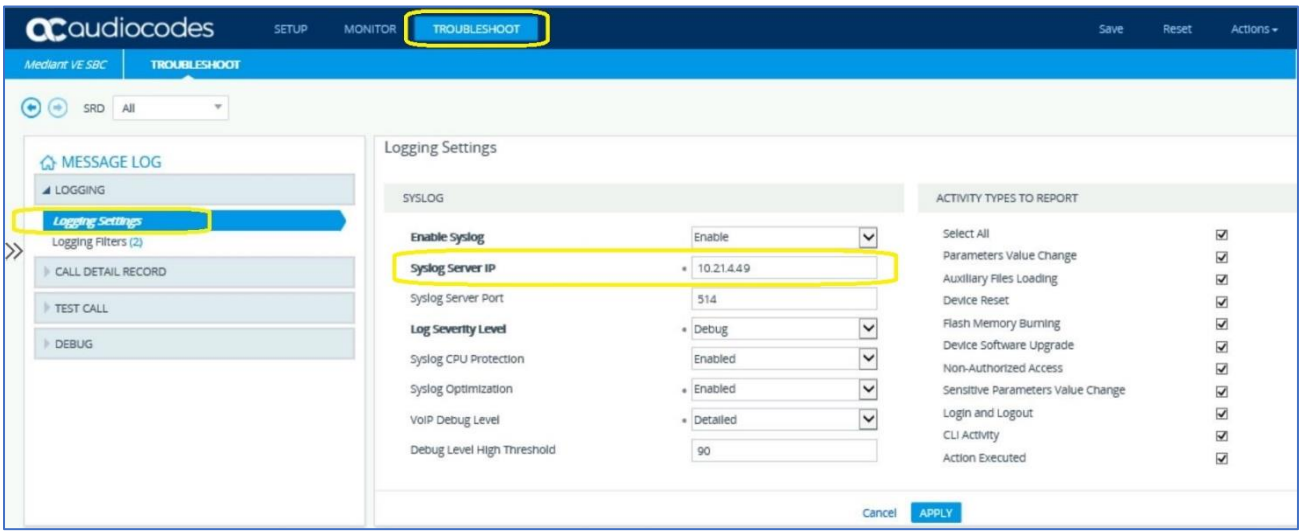

*Figure 29: SBC Configure logging to remote server.*

Collecting the audio evidence from a VoIP call requires an additional effort. First, we must configure the SBC to send the RDP traffic to the machine that we will use to collect the evidence. This can be done in the "Troubleshoot" section of the SBC by configuring the "Logging Filters" to include the "Signaling & Media & PCM" as shown in Figure 30 and in Figure 31.

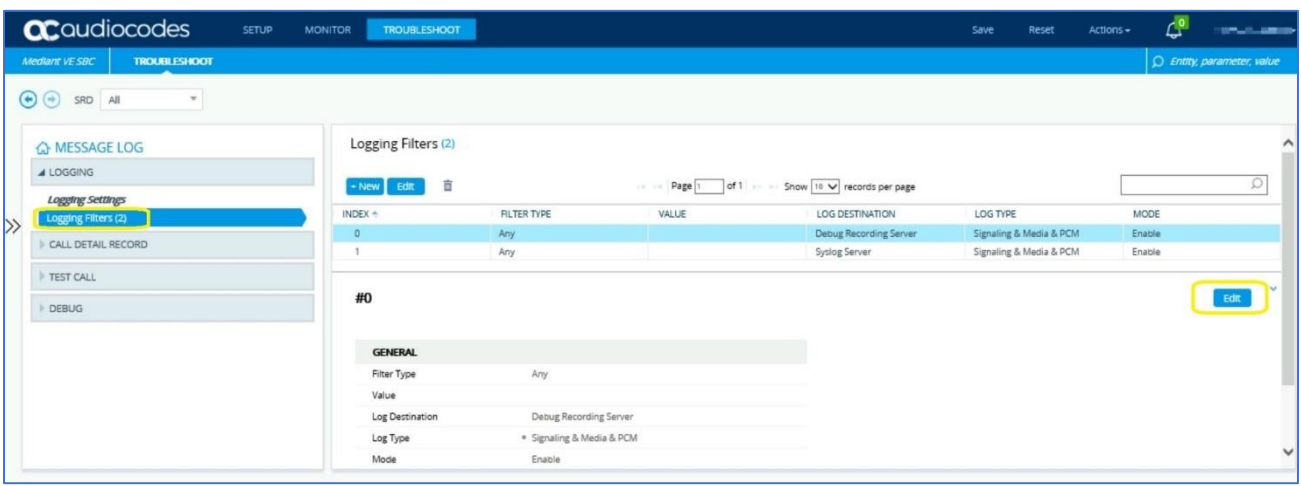

*Figure 30: SBC logging filters.*

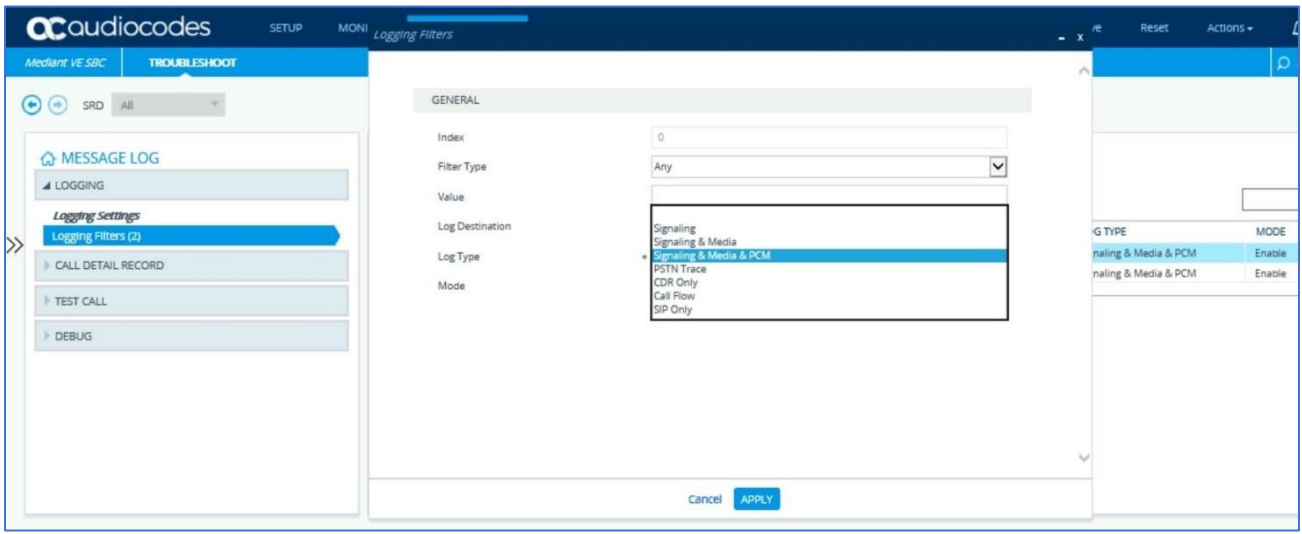

*Figure 31: SBC filter to log Media.*

Second, we need to install Wireshark in the computer which is going to receive the logs. For our tests we used Wireshark Version 3.4.6 to collect the logs and Wireshark Version 3.4.7 to analyse the logs. Usually, you should find 4 audio streams in a recording trace:

- Incoming from Teams to SBC
- Outgoing from SBC device to PSTN
- Incoming from PSTN to SBC
- Outgoing from SBC device to Teams

In Wireshark we need to open the captured trace, select "Telephony", "RTP", "RTP Streams" and for each VoIP call we will be presented with something like what is shown in Figure 32, where each of the lines is an RTP Stream.

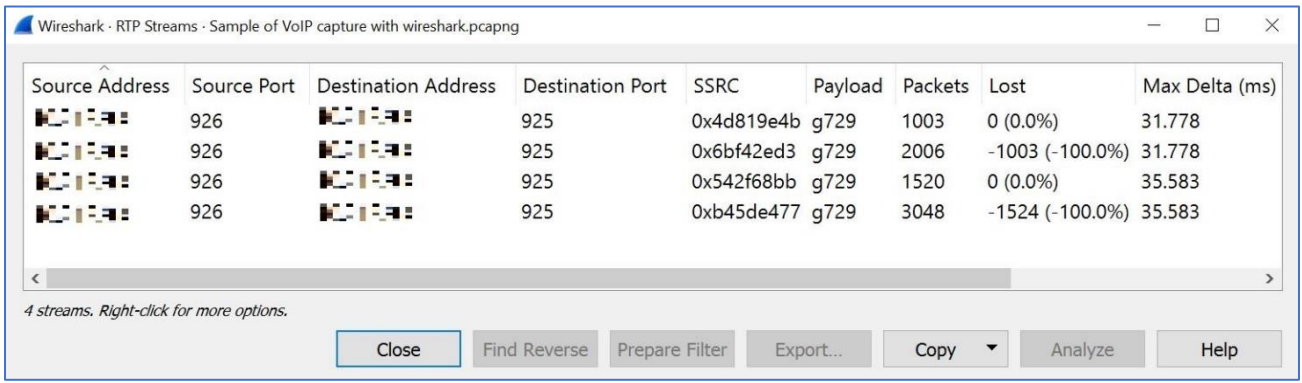

*Figure 32: RTP streams in a Wireshark VoIP capture.*

Some of the flows may have duplicated packets and in general for a forensic analysis we prefer to create a single audio file with two channels, each representing the audio of one party of the call.

Wireshark allows us to view the filters used to select each of the RTP streams as shown in Figure 33, and AudioCodes provides a useful reference guide to understand how to debug audio streams on SBC installing in Wireshark a plugin to view the debug recording packets (ACDR)[8].

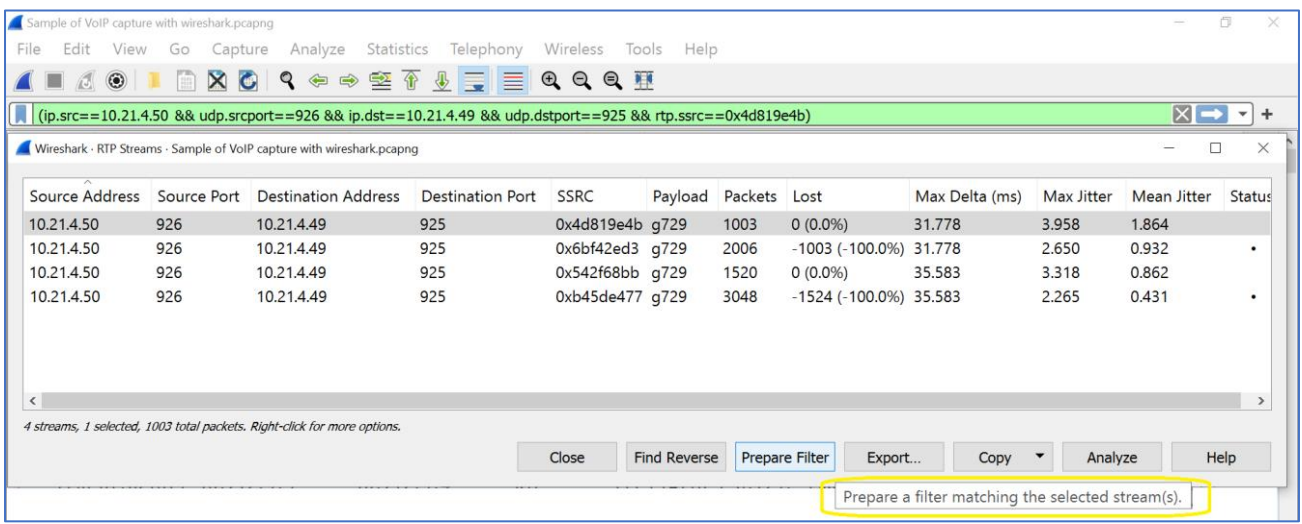

---

[8] The AudioCodes Quick Reference Guide to Debug voice with Wireshark:
https://www.audiocodes.com/media/13565/audiocodes-quick-reference-guide-how-to-debug-voice-with-wireshark.pdf

*Figure 33: Wireshark "Prepare filters" option.*

With this information, we can now elaborate our filter to select and to merge only the RTP streams we want. For example, in our scenario we achieved the result of cleaning the trace shown in Figure 17 with the following Wireshark filter: "(acdr.trace_pt == 35 and acdr.src_id == 36) or (acdr.trace_pt == 21 and acdr.src_id == 38)".

# APPENDIX C – METHODOLOGY FOR COLLECTING FORENSICS EVIDENCE FROM WALKIE TALKIE

We set up the network lab environment shown in Figure 34 to analyse with Wireshark the Wi-Fi network communication to and from a Walkie Talkie App installed on an Android mobile device.

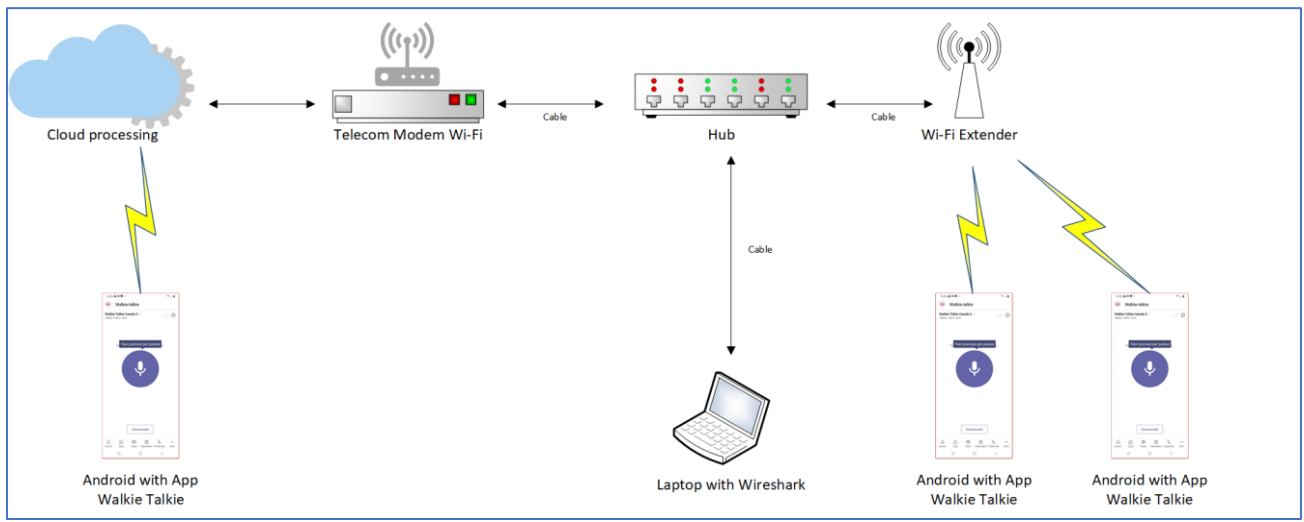

*Figure 34: Lab set up to track and analyse the traffic from Teams Walkie Talkie App*

The specification of the Android mobile device that we used to run the Walkie Talkie App that we monitored are listed in Table 5.

| Mobile device | Galaxy M21 |
|---|---|
| Android version | 11 |
| Android Kernel version | 4.14.113-20747890 #1 Tue Jan 12 22:57:51 KST 2021 |
| Android Build version | RP1A.200720.012.M215FXXU2BUAC |
| Microsoft Teams version | 1416/1.0.0.2021040701 |

| Microsoft Teams Calling Software version | 2021.06.01.30 |

*Table 5: Mobile device and Microsoft Teams mobile version*

The Android device with the Teams Walkie Talkie App that we monitored had the 192.168.1.5 IP address and the network traces show that the walkie talkie client has been communicating with several Office 365 Teams IPs in the range 52.112.0.0/14. By selecting in Wireshark "Statistics", "Conversations" and filtering for "(ip.addr == 52.112.0.0/14 && ip.addr == 192.168.1.5)" we got the traffic generated to/from the Walkie Talkie App during a Walkie Talkie session.

Figure 35 and Figure 36 show the traffic generated to/from the Walkie Talkie App during a Walkie Talkie session with another client connected to the same Team's Channel from outside the lab network. The MAC Address 80:58:f8:13:2b:5c belongs to the Android device where the App Walkie Talkie was installed, the IPs 52.114.x.x in the traces are all in the range of the IPs published by Microsoft for Office 365 Teams (Carolyn Rowe and others, 2021).

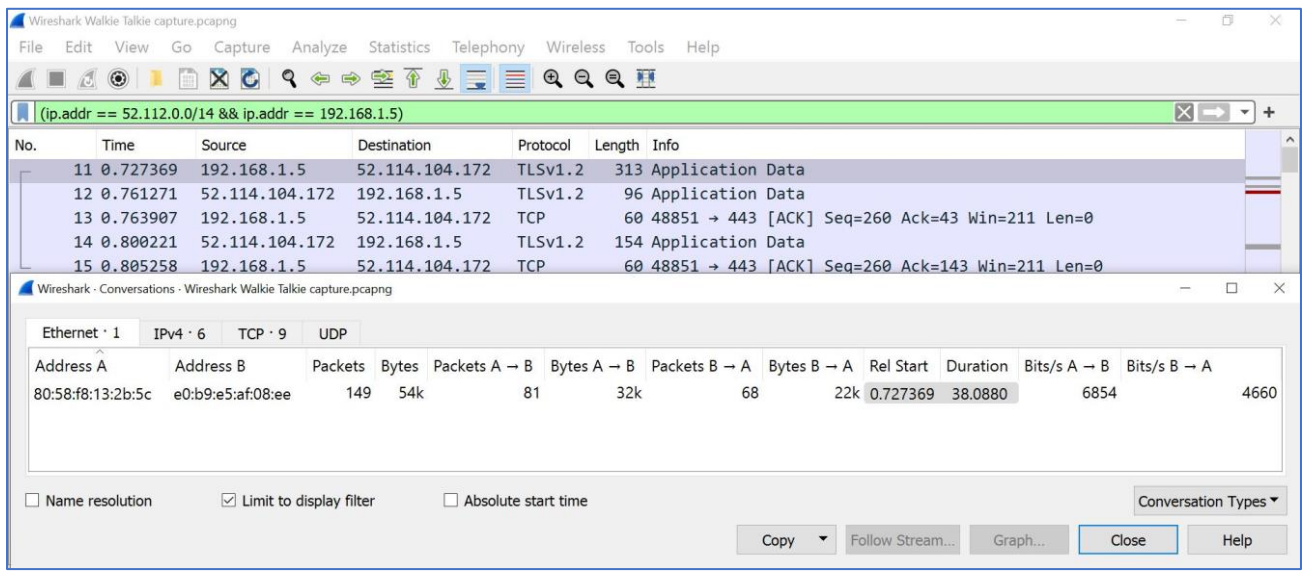

*Figure 35: MAC Addresses involved during a Walkie Talkie session*

*Figure 36: Office 365 IPs during a Walkie Talkie session[9]*

Apart from the traffic to the IPs published by Microsoft for Office 365 Teams, the Android device with the Walkie Talkie App made traffic only to Google (not relevant for our analysis) and with the local gateway IP 192.168.1.1. This information is clearly visible in Figure 37[10]. The same figure also shows the evidence of the App Walkie Talkie requesting the DNS entry for "walkietalkie.teams.microsoft.com". Figure 38 shows that a public DNS query for "walkietalkie.teams.microsoft.com" returns the IP 52.114.74.99 that is in fact the one we traced in the network traffic during a Walkie Talkie communication. Figure 39 shows that there was no SIP traffic at all in the Walkie Talkie trace.

---

[9] On Wireshark set the filter for the IPs to be visualized, then select "Statistics" and "Conversations" and flag "Limit the result to the visualizing filter."
[10] Please note the "Not" condition before the clause ip.addr = 52.112.0.0

*Figure 37: "NOT" Office 365 IPs during a Walkie Talkie session*

*Figure 38: NSLOOKUP DNS query for Walkie Talkie*

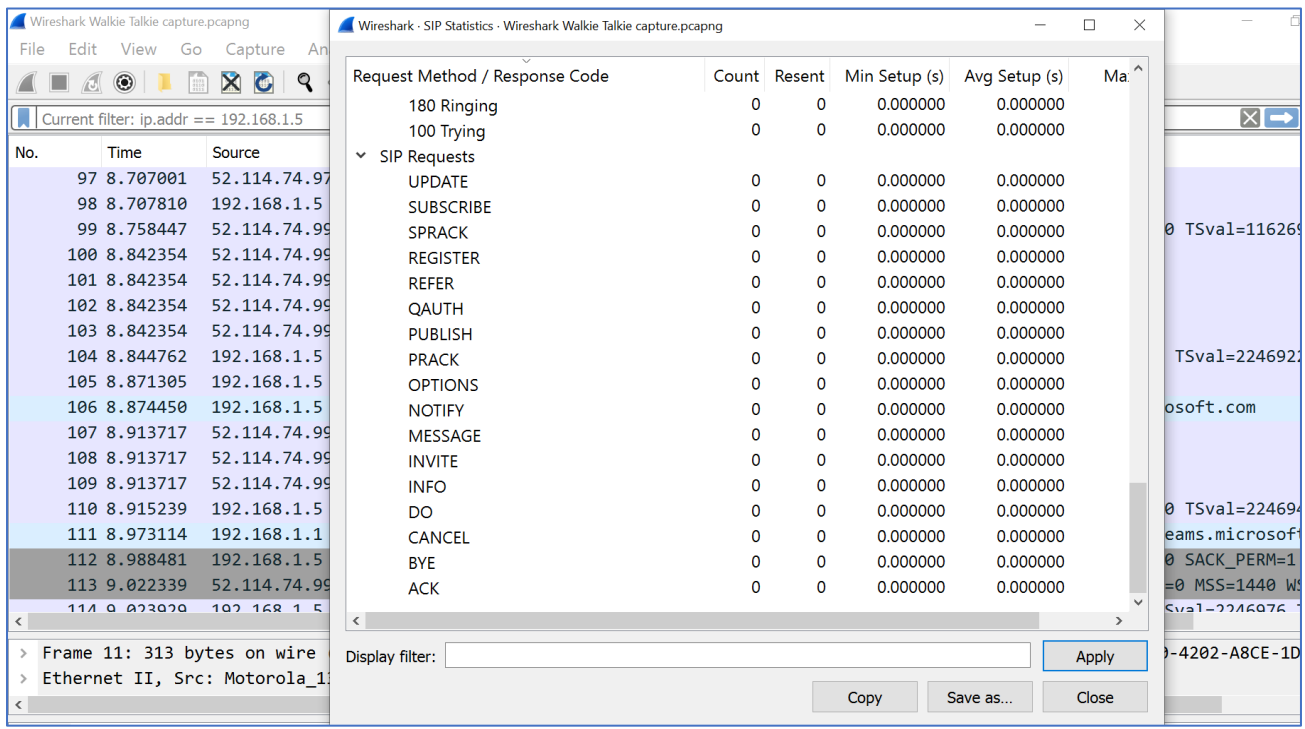

*Figure 39: No SIP traffic in a Walkie Talkie trace*

# REFERENCES

AudioCodes. (2021, 06 28). *One Voice Operations Center (OVOC)*. Retrieved from AudioCodes Ltd. Web Site: https://www.audiocodes.com/solutions-products/products/management-products-solutions/one-voice-operations-center

Carolyn Rowe and others. (2018, 08 06). *Microsoft Teams call flows.* Retrieved from docs.microsoft.com: https://docs.microsoft.com/it-it/microsoftteams/microsoft-teams-online-call-flows

Carolyn Rowe and others. (2021, 07 01). *Plan Direct Routing.* Retrieved from https://docs.microsoft.com/: https://docs.microsoft.com/en-us/microsoftteams/direct-routing-plan

Cheryl McGuire and Others. (2021, 07 12). *What is Azure Bastion?* Retrieved from https://docs.microsoft.com: https://docs.microsoft.com/en-us/azure/bastion/bastion-overview

Heidi Payne and others. (2021, 05 13). *Walkie Talkie app in Microsoft Teams.* Retrieved from https://docs.microsoft.com/: https://docs.microsoft.com/en-us/microsoftteams/walkie-talkie

jcaparas and others. (2021, 06 09). *Microsoft 365 Defender.* Retrieved from https://docs.microsoft.com/: https://docs.microsoft.com/en-us/microsoft-365/security/defender/microsoft-365-defender?view=o365-worldwide

jcaparas and others. (2021, 07 29). *Microsoft Defender for Endpoint data storage and privacy.* Retrieved from https://docs.microsoft.com: https://docs.microsoft.com/en-us/microsoft-365/security/defender-endpoint/data-storage-privacy?view=o365-worldwide

Laura Williams and others. (2021, 03 23). *Security and compliance in Microsoft Teams.* Retrieved from https://docs.microsoft.com: https://docs.microsoft.com/en-us/microsoftteams/security-compliance-overview

Mark Johnson and others. (2021, 03 17). *Assign eDiscovery permissions in the Security & Compliance Center.* Retrieved from https://docs.microsoft.com: https://docs.microsoft.com/en-us/microsoft-365/compliance/assign-ediscovery-permissions?view=o365-worldwide

Mark Johnson and others. (2021, 03 23). *Place a Microsoft Teams user or team on legal hold.* Retrieved from https://docs.microsoft.com: https://docs.microsoft.com/en-us/microsoftteams/legal-hold

Mark Johnson and others. (2021, 5 11). *Search for Teams chat data for on-premises users.* Retrieved from https://docs.microsoft.com/: https://docs.microsoft.com/en-us/microsoft-365/compliance/search-cloud-based-mailboxes-for-on-premises-users?view=o365-worldwide

Mark Johnson and others. (2021, 06 09). *Use Content search in Microsoft Teams.* Retrieved from https://docs.microsoft.com/: https://docs.microsoft.com/en-us/microsoftteams/content-search

Mike Plumley and others. (2021, 04 22). *How SharePoint Online and OneDrive for Business interact with Microsoft Teams.* Retrieved from doc.microsoft.com: https://docs.microsoft.com/en-us/microsoftteams/sharepoint-onedrive-interact#:~:text=For%20every%20user%2C%20the%20OneDrive,to%20the%20intended%20user%20only.

## BIOGRAPHICAL INFORMATION

Massimo BERNASCHI has been 10 years with IBM working in High Performance Computing. Currently he is with the National Research Council of Italy (CNR) as Chief Technology Officer of the Institute for Computing Applications. He is also an adjunct professor of Computer Science at "Sapienza" and LUISS Universities in Rome. Email: massimo.bernaschi@cnr.it

Marco NICOLETTI holds a University Degree in Computer Science and a master's degree in Computer and Information System Security. His experience includes 15 years with MICROSOFT and 12 years with PROGE-SOFTWARE. Email: nicoletti.marco@hotmail.it

## DECLARATION OF INTEREST

Declarations of interest: none.

Funding: This research did not receive any specific grant from funding agencies in the public, commercial, or not-for-profit sectors.